\documentclass{jfm}
\usepackage{graphicx}
\usepackage{amsmath}
\usepackage{float}
\usepackage[figuresleft]{rotating}
\usepackage{lscape}
\usepackage{caption}
\usepackage{epstopdf, epsfig}
\DeclareMathOperator{\e}{e}
\usepackage{xcolor}
\usepackage{color}
\usepackage{soul}
\usepackage{ulem}

\newcommand {\Derp} [2] {\frac {\partial #1} { \partial #2}}

\definecolor{brick}{rgb}{0.75,0,0}
\definecolor{rltgreen}{rgb}{0,0.5,0}
\newcommand{\revJCR}[1]{\textcolor{rltgreen}{{#1}}}

\newlength{\wdth}

\shortauthor{A. Jouin, S. Cherubini, J.-C. Robinet}

\title{Turbulent transition in a channel with superhydrophobic walls: the effect of roughness anisotropy}

\author{A. Jouin \aff{1} \aff{2},
  S. Cherubini \aff{2}
 \and J.C. Robinet\aff{1}}

\affiliation{
\aff{1} DynFluid, Ecole Nationale Supérieure des Arts et Métiers, 75013 Paris, France
\aff{2} Dipartimento di Meccanica, Matematica e Management, Politecnico di Bari, 70126 Bari, Italy
}

\begin{document}

\maketitle

\begin{abstract}
Superhydrophobic surfaces dramatically reduce skin friction of overlying liquid flows. These surfaces are complex and numerical simulations usually rely on models for reducing this complexity. One of the simplest consists in finding an equivalent boundary condition through an homogenisation procedure, which in the case of channel flow over oriented riblets, leads to the presence of a small spanwise component in the homogenized base flow velocity. This work aims at investigating the influence of such a three-dimensionality of the base flow on stability and transition in a channel with walls covered by oriented riblets. Linear stability for this base flow is investigated: a new instability region, linked to cross-flow effects, is observed. Tollmien-Schlichting waves are also retrieved but the most unstable are three-dimensional. Transient growth is also affected as oblique streaks with non-zero streamwise wavenumber become the most amplified perturbations.
When transition is induced by Tollmien-Schlichting waves, after an initial exponential growth regime, streaky structures with large spanwise wavenumber rapidly arise. Modal mechanisms appear to play a leading role in the development of these structures and a secondary stability analysis is realised to retrieve successfully some of their characteristics. The second scenario, initiated with cross-flow vortices, displays a strong influence of nonlinearities. The flow develops into large quasi spanwise-invariant structures before breaking down to turbulence. Secondary stability on the saturated cross-flow vortices sheds light on this stage of transition. In both cases, cross-flow effects dominate the flow dynamics, suggestings the need to consider these effects when modelling superhydrophobic surfaces.
\end{abstract}

\section{Introduction}

Drag is the resistance to motion experienced by a fluid flowing on a surface, generated by the difference in velocity between the solid object and the fluid. The reduction of drag represents a key factor for a whole range of physical and engineering problems involving the relative motion of a fluid on a solid surface, for instance the transport of drinkable water in pipes, of the blood in human vessels, and the motion of aircrafts in the air and of ships in the sea. For all of these applications, one of the main sources of drag is the skin friction between the molecules of the fluid and the solid surface over which it flows, whose magnitude  depends on the properties of both the surface and the fluid flowing on it. In the vicinity of a smooth hydrophilic surface, the flow must decelerate until reaching zero velocity at the wall, self-inducing a strong resistance to motion due to skin friction. This resistance to motion can be reduced by  decreasing the gradient of velocity between the surface and the flow itself, which can be achieved by using particular surfaces or coatings, able to assure shear-free or slip wall conditions. One example of this kind of surfaces is given by superhydrophobic (SH) surfaces \citep{Rothstein2010}. The superhydrophobicity of a surface is due to its nanostructure, which is composed by a hierarchical structure of microroughnesses 
that trap the air underneath them, reducing the surface of contact with the water droplets and the wetting of the surface. Such a hierarchical, rough nanostructure can be found on many biological surfaces such as those of lotus leaves, butterfly wings, duck feathers and water striders legs, and have inspired the engineering of biomimetic non-wettable materials for applications that range from self-cleaning to anti-icing. 
\\
Recently, some experiments and numerical simulations of flows on SH channels \citep{Lee2016} have revealed that these surfaces are also capable of considerably reducing the drag in both laminar and turbulent conditions.
Using different types of micro-posts and micro-ridges, \cite{ou_laminar_2004} have been able to show that an increasing shear-free area induced by carefully engineered micro-roughnesses induces an increased slip length and a consequent drag reduction for laminar channel flows. Along the air-water interfaces, the authors measured slip velocities larger than the $60\%$ of the averaged velocity, corresponding to a drag reduction larger than $40\%$. More experiments have been performed using superhydrophobic surfaces characterized by nano-sized structures such as ridges or needles \citep{Choi2006}, trying to engineer the nano-structures able to maximize the attainable drag reduction \citep{Lee2009}. In particular, \cite{Steinberger2007} have found that microposts are less effective in reducing drag than ridges, since the flow has to decelerate and accelerate between different posts resulting in a lower slip length. Direct numerical simulations have followed these experiments; in order to model the interaction between air pockets and water, which might be very complex, the air-water interface has been assumed flat and the viscosity of the air trapped in the micro-ridges has been neglected \citep{Ybert2007}. Direct numerical simulations have been also performed for turbulent flows, first by imposing a slip condition using an arbitrary slip length \citep{Min2004},
and then by alternating shear-free and no-slip boundary-conditions on a flat surface \citep{martell_perot_rothstein_2009, Martell2010}. 
Comparison of these alternative approaches have shown that using spatially-homogeneous partial slip boundary conditions is a reliable approach in SH surfaces modelling: it is capable of correctly predict properties for both laminar \citep{choi_apparent_2003,ou_laminar_2004} and turbulent regimes \citep{zhang_mechanisms_2015,zhang_drag_2016}. Stability thresholds and instability mechanisms are also accurately retrieved \citep{yu_linear_2016}. Still, the accurate description of SH surfaces rests on solving two fundamental difficulties: finding an adequate slip length modelling the effect of the SH wall and ensuring that this slip length describes a wetting-stable configuration within the micro-roughnesses (namely, one able to retain the trapped lubricant even in turbulent conditions). 
The stability of an air-water interface in the context of SH coatings is complex and based on several phenomena, as studied parametrically by \citet{seo_turbulent_2018}. Wetting transition, that causes plastron depletion and ultimately leads to augmented drag,  may be caused by capillarity or stagnation pressure effects. Recently, \cite{Seo2017} have shown that for limited values of the Weber number and of the roughness size, wetting stable conditions are maintained, while choosing larger values of these parameters may lead to destabilization of the liquid-gas interface.
Considering the value of the slip length required for accurately  modelling a SH surface, once chosen a type of roughness and a set of parameters ensuring wetting-stable conditions, one can use a homogenisation approach such as that proposed by  \cite{bottaro_effective_2020} for retrieving a slip length value physically representative of that macroscopically obtained by SH microroughnesses of different shapes and size (for a review, see \cite{bottaro_2019}). 
Provided that these two conditions (wetting stability and adequate slip lengths) are satisfied, the use of spatially-homogeneous partial slip boundary conditions is appropriate for simulating the effect of SH surfaces on the flow dynamics. 
\\
\cite{Park2013} have found that the potential drag reduction of SH coatings is much larger in turbulent flows than in laminar ones, and this effect might be due to the damping of wall-turbulence induced by the presence of a slip length. In fact, this increased effectiveness in drag reduction appears to be mostly  due to the weakening of the coherent structures sustaining turbulence, rather than by the slip condition itself.  This indicates that superhydrophobic surfaces might be also efficient in delaying transition to turbulence; however,  the effect of those kind of surfaces on turbulent transition is a point which has been still not widely investigated. In general, transition from laminar to turbulent flow induces a strong increase in skin friction, along with a strong increase of the drag. Thus, in the transitional regime, the competition between drag decrease due to the surface micro structure, and drag increase due to transition to turbulence might provide surprising results. 
Up to now, very few studies have been performed on transition to turbulence of a laminar flow over SH surfaces, focusing at first on the very first phase of transition, namely, linear instability \citep{LaugaCossu2005,min_effects_2005}. For a channel flow, it has been proved by a local instability analysis that, when imposing a simple slip condition, the onset of two-dimensional Tollmien-Schlichting waves is considerably postponed, allowing the flow to stay laminar up to larger Reynolds numbers, and further decreasing the drag. However, shear flows very often experience subcritical transition to turbulence, due to the transient growth of non-modal disturbances, bypassing the asymptotic growth of Tollmien-Schlichting waves \citep{schmid_stability_2001,Farano_2016}. 
 In particular, it has been shown in \cite{min_effects_2005} that slip boundary conditions have a strong influence on the linear growth of modal disturbances, but a very weak influence on the maximum transient energy growth of perturbations at subcritical Reynolds numbers, concluding that slip boundary conditions are not likely to have a significant effect on the transition to turbulence in channel flows. 
 By means of global stability analyses, \cite{tomlinson_papageorgiou_2022} have shown that in the presence of SH grooves, additional instabilities may arise, with critical Reynolds numbers small enough to be achievable in applications.
 \\
 The validity of linear stability results on laminar-turbulent transition itself, which is an intrinsically nonlinear phenomenon, has been recently assessed by \cite{picella_laminarturbulent_2019,picella_influence_2020}, who have confirmed that SH surfaces strongly influence transition induced by wall-close disturbances, such as TS waves, even at subcritical Reynolds number, but have a weak effect on the subcritical  growth of coherent structures lying farther from the wall, such as streaks and streamwise vortices. \cite{Cherubini2021} reported a strong effect of boundary slip on the transient growth of nonlinear optimal  perturbations: in particular, while the maximal energy growth is considerably decreased, the shape of the optimal perturbation barely changes,
indicating the robustness of optimal perturbations with respect to wall slip.
A~recent study by \cite{davis_park_2020} has assessed the influence of slip at the wall on nonlinear traveling waves solutions of the Navier--Stokes equations. Depending on their structure, the transition to turbulence of these nonlinear solutions is delayed or advanced. This indicates that SH surfaces considerably affect energy growth mechanisms relying mostly on nonlinearity, so that linear stability results are not always indicative of the effectiveness of these surfaces on delaying transition. 
Moreover, to the best of the authors' knowledge, practically all literature studies dealing with the effect of SH surfaces on turbulent transition, consider the case of statistically isotropic micro-roughnesses, therefore allowing for the use of a single slip length, $L_s$, acting on both wall-parallel directions. 
On the other hand, it has been shown by \cite{Min2004} that while streamwise slip condition induces a strong drag decrease, a condition of spanwise slip increases the drag. These results show that the amount of slip provided by these surfaces in the streamwise and spanwise directions has a non trivial effect on turbulence and transition, which is a point that deserves more accurate investigations.  A number of studies have taken into account the case of anisotropic roughnesses, using different slip lengths in the streamwise and spanwise direction \citep{Min2004,KhoshAghdam2016,pralits_stability_2017}, although only \cite{min_effects_2005} have focused on turbulent transition, finding that spanwise slip case induce an earlier transition while streamwise slip considerably delays it. However, this study considered a single value of the slip length for both the streamwise and spanwise directions, which was then nullified on one of the two wall-parallel directions, while not considering the case of different (non-zero) values of the slip. Whereas, in the case of oriented (non-isotropic) micro-posts, the slip length assumes different values in the streamwise and spanwise directions, since the homogenized boundary condition for the wall velocities assume a tensorial form \citep{pralits_stability_2017}. The effect on turbulent transition of such tensorial slip conditions, which are representative of non-isotropic SH micro-roughnesses, remains entirely to be unveiled. 
\\
In order to investigate this issue, in the present paper we consider the case of oriented superhydrophobic riblets-like micro-roughnesses \citep{pralits_stability_2017}, and investigate their impact on modal transition. It will be shown that the non-isotropicity of the roughnesses gives origin to a new modal instability, similar to cross-flow instabilities recovered on the flow over swept-wings, and will considerably alter nonmodal stability as well. The transition to turbulence originated by the former instability, as well as by the slip-modified TS waves, will be studied in detail using direct numerical simulations and secondary stability analyses. 
\\
The paper is structured as follows. In section \S\ref{sec:PROBLEM_FORMULATION} we present the governing equations and the methods used to implement  the SH surface. In section \S\ref{sec:LSA} we show, using stability and transient growth analysis, how the dynamics of infinitesimal perturbations is influenced by the considered SH surfaces. Section \S\ref{sec:DNS} reports the results of direct numerical simulations of laminar-turbulent transition, for the different considered instabilities.  
A final discussion and conclusions are given in section \S\ref{sec:Conclusions}.

\section{Governing equations}\label{sec:PROBLEM_FORMULATION}

\begin{figure}
    \centering
    \includegraphics[width=1.\textwidth]{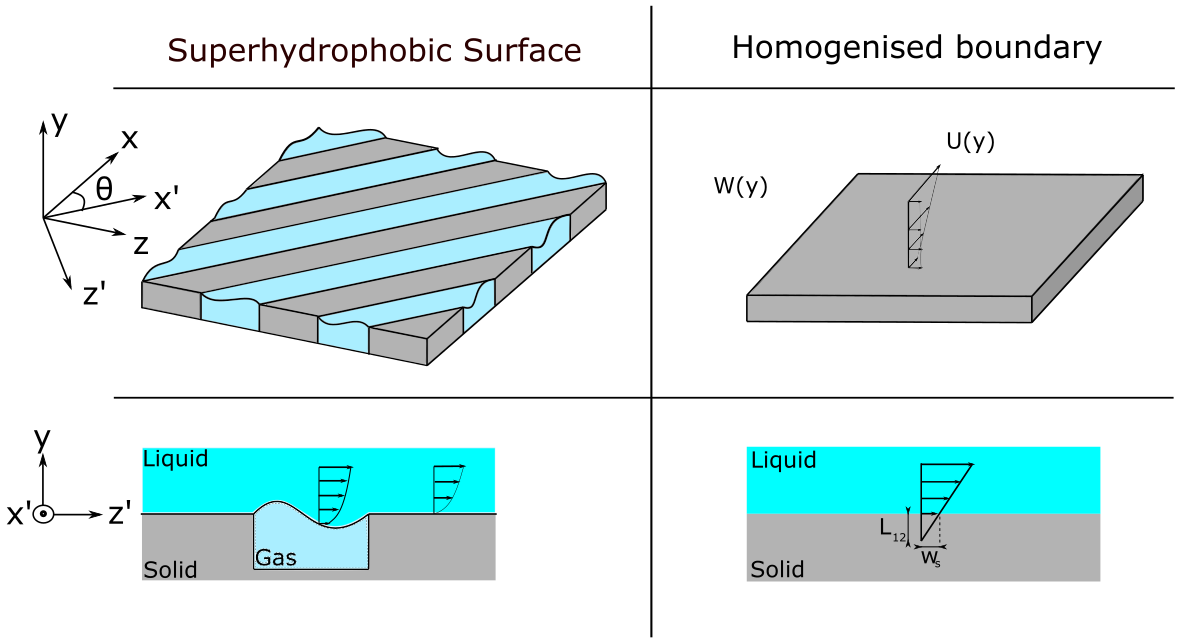}
    \caption{Channel flow over superhydrophobic surfaces oriented at an angle $\theta$ with respect to the streamwise direction. The superhydrophobic surface depicted in the left is homogenised and reduces to an anisotropic boundary condition.}
    \label{fig:Schema}
\end{figure}

The flow of an incompressible Newtonian fluid in a channel of height $2h$ with superhydrophobic (SH) walls is considered. The reference frame is chosen as $({x},{y},{z})$, with $x$ being the streamwise direction, $y$ the wall normal and $z$ the spanwise direction. Introducing the reference velocity $U_r = 3U_a/2$, with $U_a$ being the average of the base flow velocity $U_a = 1/2h \int U dy$
and the kinematic viscosity of the fluid $\nu$, the Reynolds number is defined as $Re = U_rh/\nu$. The dynamics of the flow is governed by the Navier-Stokes equations:

\begin{align}
    \Derp{\mathbf{U}}{t} = - (\mathbf{U} \cdot \mathbf{\nabla}) \mathbf{U} &-\mathbf{\nabla} p +\frac{1}{Re}\mathbf{\nabla}^2 \mathbf{U} \\
    \mathbf{\nabla} \cdot \mathbf{U} &= 0,
    \label{NS}
\end{align}
where $\mathbf{U}=(U,V,W)^T$ is the velocity vector and $p$ is the pressure. The flow is periodic in the streamwise and spanwise directions, and different domain sizes will be considered, as detailed in section \ref{sec:DNS}.
The walls of the channel are covered with SH riblet-like roughnesses oriented with an angle $\theta$, defined with respect to the $x$ direction (see figure \ref{fig:Schema}).

For $\theta = 0^o$, the grooves are longitudinal while $\theta=90^o$ corresponds to transverse riblets. The effect of longitudinal SH riblet-like roughnesses can be modelled using equivalent streamwise and spanwise slip lengths $\lambda_{\parallel}$ and $\lambda_{\perp}$ \citep{gogte_effective_2005,belyaev_effective_2010}, which lead to the homogenised boundary conditions $U=\lambda_{\parallel}\partial_y U$, $V=0$, and $W=\lambda_{\perp}\partial_y W$. When the grooves are aligned with the mean pressure gradient, the streamwise slip length is twice the spanwise one, i.e. $\lambda_{\parallel} = 2 \lambda_{\perp}$ \citep{philip_flows_1972, asmolov_effective_2012}. Whereas, when the riblets are not aligned with the pressure gradient, 
 the homogenised boundary conditions for the Navier-Stokes equations have a more general form:

\begin{equation}
\begin{bmatrix}
U \\
W
\end{bmatrix} = \mp \mathbf{L} \partial_y
\begin{bmatrix}
U \\
W
\end{bmatrix}, \quad V = 0, \quad \text{at} \quad y=\pm 1
\label{BC}
\end{equation}
where the mobility tensor $\mathbf{L}$ depends on $\lambda_{\parallel}$, $\lambda_{\perp}$ and on  the rotation matrix $\mathbf{R}(\theta)$,  allowing the rotation of the roughness of an angle $\theta$ \citep{pralits_stability_2017}: 
\begin{equation}
    \mathbf{L} = \mathbf{R} 
    \begin{pmatrix}
    \lambda_{||} & 0 \\
     0 & \lambda_{\perp}
    \end{pmatrix}
    \mathbf{R}^T=
    \frac{\lambda_{\parallel}}{2}
    \begin{pmatrix}
    1 + \cos^2 \theta & \cos{\theta}\sin\theta  \\
    \cos{\theta}\sin\theta & 1 + \sin^2 \theta.
    \end{pmatrix}
\end{equation}
Notice that the mobility tensor can be expressed as a function of $\lambda_{\parallel}$ and $\theta$ only. 
\\
Following \cite{pralits_stability_2017}, we set $\lambda_{\parallel}=0.03$ and $\theta=45^o$ yielding $L_{11} = L_{22} = 0.0225$ and $L_{12} = L_{21} = 0.0075$. These effective slip lengths comply with the wetting-stable conditions reported by \citet{seo_turbulent_2018}.
As found by \cite{pralits_stability_2017}, the laminar base flow takes the form $\mathbf{U}_0 = (U_0(y),0,W_0)$, with:
\begin{equation}
    U_0 = \frac{ 2L_{11} + 1 - y^2}{1 + 3L_{11}}, \qquad W_0 = \frac{2L_{12}}{1 + 3L_{11}} 
    \label{BF}
\end{equation}

The streamwise component of the base flow is similar to that found for a Poiseuille flow with a partially slippery wall with $\lambda_{\parallel} = \lambda_{\perp}$  \citep{philip_flows_1972, picella_laminarturbulent_2019}. Note that $U_0$ has the same mass flux as a plane Poiseuille flow with no slip. The Reynolds number remains fixed when the slip length is changed, justifying the use of $U_r$ as the reference velocity. The presence of a spanwise  base flow component, which remains constant across the wall-normal direction may seem counter-intuitive. However, being one order of magnitude smaller than the streamwise component, the spanwise one will mostly affect the near-walls regions, while having negligible effects in the centre of the channel.  

\section{Linear stability analysis}\label{sec:LSA}

The instantaneous flow field is now decomposed as the sum of the previously described steady base flow and an unsteady disturbance having small amplitude. Introducing the state vector $\mathbf{Q}(\mathbf{x},t) = [U,V,W,P]^T $, we have $\mathbf{Q}(\mathbf{x},t) = \mathbf{Q}_0(\mathbf{x}) + \epsilon \mathbf{q}(\mathbf{x},t)$, with $\epsilon << 1$. Navier-Stokes equations are then linearised with respect to the base flow $\mathbf{Q}_0=[\mathbf{U}_0,P_0]^T$, yielding the following system of equations:
\begin{align}
    \Derp{\mathbf{u}}{t} + (\mathbf{U}_0\cdot \nabla)\mathbf{u} + (\mathbf{u}\cdot \nabla)\mathbf{U}_0 &= -\nabla p + \frac{1}{Re}\nabla^2 \mathbf{u}, \label{EqLin1}\\
    \nabla \cdot \mathbf{u} &= 0
    \label{EqLin2}
\end{align}

together with the boundary conditions for the velocity perturbations:
\begin{equation}
\begin{bmatrix}
u \\
w
\end{bmatrix} = \mp \mathbf{L} \Derp{}{y}
\begin{bmatrix}
u \\
w
\end{bmatrix}, \quad v = 0 \quad \text{at} \quad y=\pm 1
\label{BC}
\end{equation}

\subsection{Modal stability analysis}

The system being periodic in the streamwise and spanwise directions, the perturbation state vector $\mathbf{q}$ can be expanded in normal modes such that
\begin{equation}
    \mathbf{q}(\mathbf{x},t) = \mathbf{\hat{q}}(y)\exp{[i(\alpha x + \beta z  - \omega t)]} + c.c.
    \label{ModalExpansion}
\end{equation}

with c.c. the complex conjugate, $\alpha$, $\beta$ being respectively the streamwise and spanwise wave numbers and $\omega$ an angular frequency. Temporal stability is investigated, implying $\alpha$ and $\beta$ are real while $\omega = \omega_r + i\omega_i$ is a complex number whose imaginary part determines the asymptotic stability of the base flow $\mathbf{U}_0$ for a given mode $(\alpha, \beta)$. Thus, substituting (\ref{ModalExpansion}) into the linearised Navier-Stokes equations (\ref{EqLin1})-(\ref{EqLin2}) results in a generalised eigenvalue problem which can be solved by means of a spectral collocation code \citep{trefethen_spectral_2000, schmid_analysis_2014}.
\begin{figure}
    \centering
    \includegraphics[width=1.\textwidth]{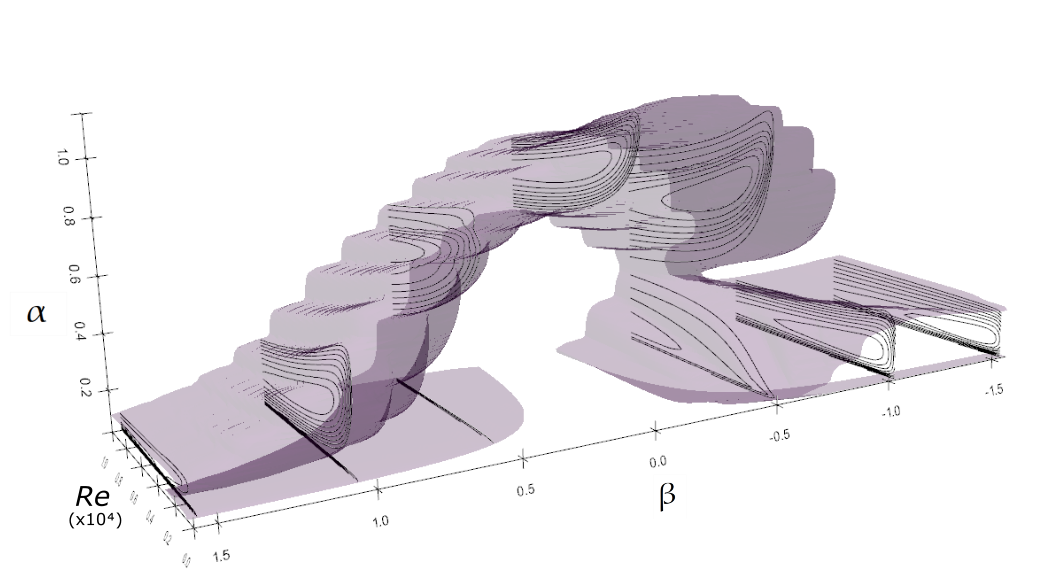}\\
    \includegraphics[width=0.45\textwidth]{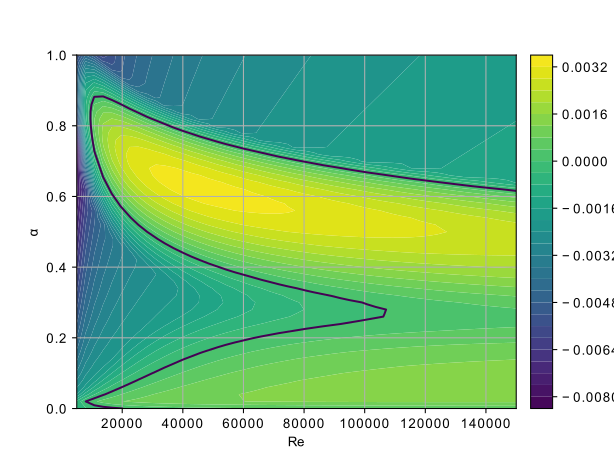}
    \includegraphics[width=0.45\textwidth]{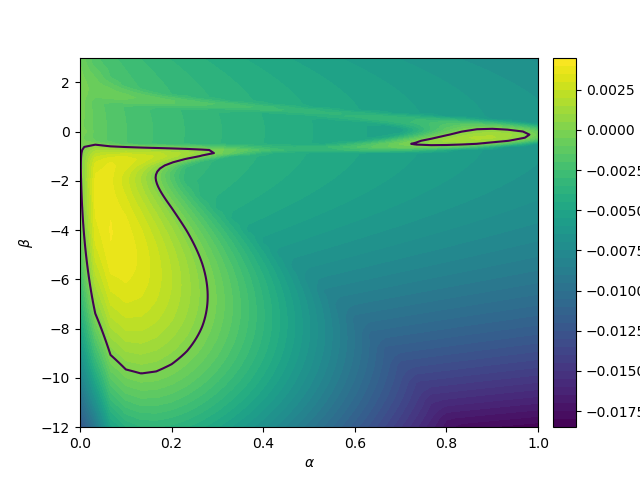}
    \caption{Top: Neutral surface in the $\alpha-\beta-Re$ space. Contour plots: slices of the isosurface for a few chosen $\beta$. Isocontours are only depicted in the unstable regions ($\omega_i > 0$). Bottom: Neutral curve in the $\alpha-Re$ plane for $\beta=-0.5$ (left) and in the $\alpha-\beta$ plane for $Re=12000$ (right). Black contours correspond to the projection of the neutral isosurface in the respective planes.}
    \label{fig:Neutral_Cuts}
\end{figure}
In the present configuration, Squire's theorem does not hold as shown by \citet{pralits_stability_2017}.

The stability of the flow is investigated in the full $(\alpha, \beta, Re)$ domain,  for $\lambda_{\parallel} = 0.03$ and $\theta=45^o$. For a parametric study over the influence of $\lambda_{\parallel}$ and $\theta$ on the linear stability analysis, the reader is referred to \citet{pralits_stability_2017}.
Mainly, two main unstable zones can be identified: a patch for small values of $\alpha$ and  $\beta \ne 0$ and a horseshoe region at small $\alpha$ and $\beta$ closer to zero. A strong asymmetry of the neutral isosurface can be seen: perturbations are most unstable in the opposite direction of the cross-flow base flow velocity ($\beta < 0$). This has already been noticed for swept wings \citep{mack_boundary-layer_1984}. Slices of the neutral isosurface in the $\alpha-Re$ and $\alpha-\beta$ planes, showing the regions of strongest instability, are provided  in the bottom of Figure \ref{fig:Neutral_Cuts}.
\begin{figure}
    \centering
    \includegraphics[width=0.45\textwidth]{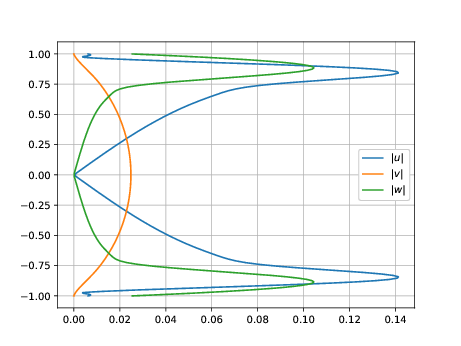}
    \includegraphics[width=0.45\textwidth]{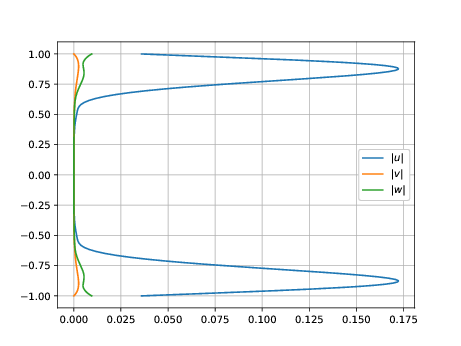}
    \includegraphics[width=0.45\textwidth]{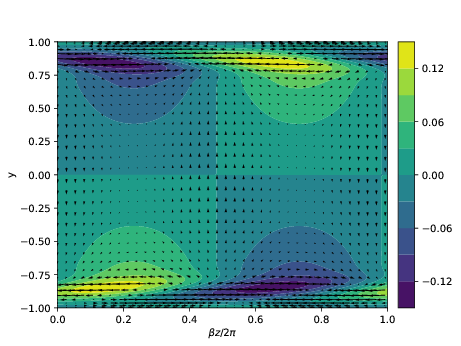}
    \includegraphics[width=0.45\textwidth]{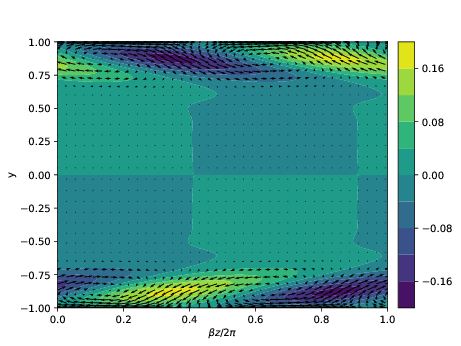}
    \caption{Eigenfunctions of the most unstable mode for the two identified regions of interest: for the left column, $\alpha = 0.7$, $\beta=-0.6$ and $\omega = 0.0004-0.1643j$ (TS wave) and for the right column, $\alpha = 0.2$, $\beta=-6$ and $\omega = 0.0033+0.02245j$ (CF vortices). Top row: absolute value of the disturbance velocity components. Bottom row: the contour plot represents the streamwise velocity disturbance while quiver plot shows the $v-w$ cross-flow.}
    \label{fig:ModesLinear}
\end{figure}

For further insight, Figure \ref{fig:ModesLinear} displays several representations of the eigenfunctions of the most unstable mode for both regions. The horseshoe instability region is reminiscent of the unstable region of both classic and homogeneous slip Poiseuille flows. The usual shape of a 3D TS-wave \citep{zang_numerical_1989} is also retrieved. The second region exhibits modes with tilted counter-rotating vortices. These vortices are quasi-stationary ($\omega_i = 0.022$) and propagate almost perpendicularly to the streamwise direction ($\phi = \arctan(\beta_0/\alpha_0 \approx 88^o$).
All these characteristics are reminiscent of unstable modes of swept boundary layers \citep{mack_boundary-layer_1984, saric_stability_2003, serpieri_three-dimensional_2016} or rotating disks flows \citep{lingwood_absolute_1995,imayama_laminarturbulent_2014} linking this region to cross-flow (CF) related instability mechanisms. This is unusual: cross-flow instability is usually depicted as a mainly inviscid mechanism \citep{saric_stability_2003}, thus necessitating an inflection point in the spanwise velocity $W_0$. For a fixed Reynolds number, cross-flow related modes are more unstable than the Tomllmien-Schlichting (TS) waves. The presence of this new instability region can be explained in a rather simple way. The boundary conditions (\ref{BC}) represent a system of equations for $(u,w)$ which, assuming 1D profiles, can be solved. The solution is physically meaningful only in the close vicinity of the walls but can still provide some interesting insight on the physics in this region. Thus, for a wall at $y=0$, the inversion of the system (\ref{BC}) yields:

\begin{equation}
\begin{bmatrix}
u \\
w
\end{bmatrix}(y) =
\begin{bmatrix}
U_0^+ \\
U_0^+
\end{bmatrix}
\exp{(y/L^+)} +
\begin{bmatrix}
U_0^- \\
-U_0^-
\end{bmatrix}
\exp{(y/L^-)}
\end{equation}
 
where $L^+ = L_{11} + L_{12}$, $U_0^+ = \frac{1}{2}(u_0 + w_0)$, $L^- = L_{11} - L_{12}$ and $U_0^- = \frac{1}{2}(u_0 - w_0)$, while $u_0$ and $w_0$ are the values of the $u$ and $w$ perturbations at the wall.

Taking advantage of the modal assumption (\ref{ModalExpansion}), the normal velocity component $v(y)$ can be deduced from the continuity equation and reads:

\begin{equation}
    v(y) = -i\left[ (\alpha + \beta)U_0^+L^+\exp{(y/L^+)} + (\alpha - \beta) U_0^-L^-\exp{(y/L^-)}  \right]
\end{equation}

Recalling that the normal velocity is zero at $y=0$, a relation linking the different parameters can be obtained: 

\begin{equation}
    \alpha = \revJCR{-} \frac{U_0^+L^+ - U_0^-L^- }{U_0^+L^+ + U_0^-L^-} \beta. \quad
    \label{EqWall}
\end{equation}

From (\ref{EqWall}), two extreme cases can be distinguished: a first one in which $w_0 >> 0$ and a second one with $w_0 \approx 0$. Since $L^+$ and $L^-$ are fixed parameters, the former leads to $U_0^+L^+ >> U_0^-L^- $ and $\alpha \approx \revJCR{-}\beta$ while the latter implies $U_0^+L^ + \approx U_0^-L^- $ and $\beta >> \alpha$. The boundary condition appears to act as a mode selector, and these two distinct behaviours  roughly correspond to the previously identified instability regions.

Despite this model, the influence of the boundary condition on these two instability mechanisms is not completely clear. Since TS waves are viscosity-induced near-wall structures, they will be affected by a decrease of the velocity gradients near the walls caused by the slip boundary conditions \citep{min_effects_2005, picella_laminarturbulent_2019}. However, cross-flow effects are also prominent in the vicinity of the walls and a strong influence on the evolution of the 3D TS wave is expected. Regarding cross-flow modes, they are characterised by counter-rotating vortices which, despite being rather weak, affect the stability of the flow by redistributing low-momentum flow from the near-wall regions to the centre of the channel and, conversely, re-injecting high-momentum flow near the walls \citep{bippes_basic_1999, serpieri_cross-flow_2018}, potentially destabilising the flow. Once again, the intensity reduction of the velocity gradients near the walls will impact the strength of the shear layers and the redistribution mechanism, albeit it is still not clear to which extent.

\subsection{Non-modal stability analysis}

In subcritical conditions, non-modal mechanisms, linked to the non-normality of the Navier-Stokes equations, can induce a transient growth of the energy of small perturbations. Such a growth can be several orders of magnitude higher than the initial energy perturbation \citep{gustavsson_energy_1991, reddy_energy_1993} thus playing a crucial role in the transition to turbulence. This behaviour is typically investigated by maximising the finite-time amplification of an initial velocity perturbation $\mathbf{u}_0$ \citep{butler_threedimensional_1992}. Mathematically, the quantity to be optimised can be written as:
$$ G(T) = \max_{\mathbf{u}_0} \frac{E(\mathbf{u}(T))}{E(\mathbf{u}_0)}$$ 
where $E(T)$ is the kinetic energy of a perturbation at target time $T$, defined as $$ E(\mathbf{u}(t)) = \frac{1}{2L_xL_z} \int_V |\mathbf{u}|^2 dV, $$
where V is the volume of the computational domain of streamwise and spanwise length $L_x$ and $L_z$, respectively.

\begin{figure}
    \centering
    \includegraphics[scale=0.6]{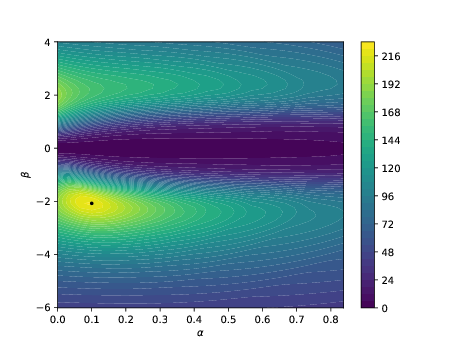}
    \caption{Optimal energy gain contours in the $\alpha-\beta$ plane for $Re=1000$. The Reynolds number is chosen sufficiently low to ensure stability of the flow. The black dot corresponds to the maximum energy gain obtained for $\alpha=0.1$ and $\beta = -2.0$.}
    \label{fig:TG}
\end{figure}

Figure \ref{fig:TG} provides the optimal energy gain $G$ in the $\alpha-\beta$ plane, which appears to be strongly affected by the presence of the wall slip. The maximum gain can be obtained for a small but non-zero $\alpha$ and for $\beta=-2$, differently from the channel flow with no-slip walls \citep{butler_threedimensional_1992}. Previous studies \citep{min_effects_2005, picella_laminarturbulent_2019} have assessed that isotropic slip conditions have a marginal effect on transient growth. Thus, we conjecture that transient growth is mainly affected by cross-flow related effects. Notice that the region of maximum energy growth overlaps with the cross-flow instability region, hinting at the coexistence of both mechanisms. Similar results were found in the non-modal analysis conducted by \citet{breuer_transient_1994} and \citet{corbett_optimal_2001} on swept wings.

\begin{figure}
    \centering
    \includegraphics[width=0.45\textwidth]{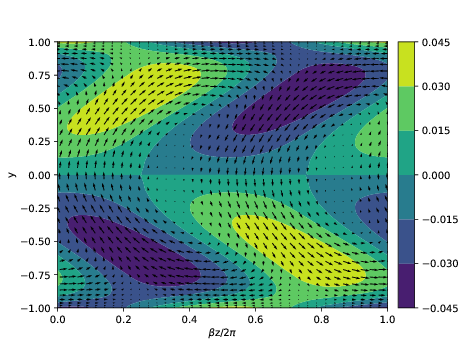}
    \includegraphics[width=0.45\textwidth]{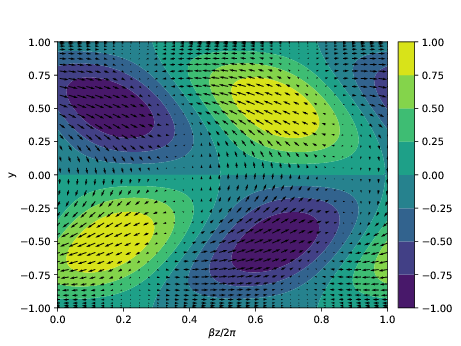}
    \caption{Optimal perturbation at time $t = 0$ (left) and at target time $t=T = 82$ (right). The contour plot represents the streamwise component of the perturbed velocity while the quiver plot depicts the $v-w$ cross-flow.}
    \label{fig:OptPert}
\end{figure}

The optimal perturbation ensuring the maximum transient growth is given in Figure \ref{fig:OptPert}. The initial perturbation (left frame) consists in quasi-streamwise tilted counter-rotating vortices. The vortices are tilted in the opposite direction of the cross-flow component of the base flow and distorted by the slip conditions in the near-wall regions. The cross-flow components have an amplitude one order of magnitude higher than the streamwise component. At the optimal time, the perturbation takes the form of velocity streaks slightly tilted in the cross-flow direction. Non-modal growth is usually the product of two mechanisms: the Orr mechanism and the lift-up effect \citep{brandt_lift-up_2014}. Since the Orr mechanism is more efficient for short ($\mathcal{O}(10)$) time scales and high streamwise wavenumbers \citep{butler_threedimensional_1992}, in the present case the transient growth is most probably linked to the lift-up effect \citep{landahl_note_1980}. However, differently from the classical case of the channel flow with no-slip walls, in the present case the perturbation can transport both the streamwise and spanwise base flow shear, inducing changes in both the optimal perturbations and the optimal gain. 
Notice that the shape of this optimal perturbation resembles that of weakly nonlinear optimal perturbations computed for a channel flow by \cite{Pralits2015}.

\citet{ellingsen_stability_1975} derived a model which showed algebraic growth for the amplitude of streamwise independent perturbations in time. A generalisation of this model for 3D base flows is not trivial, except for small angles for which the mechanisms remain similar. However, in the specific case of a constant spanwise component of the base flow, an extension of this model is possible and has been derived in Appendix \ref{appendix:liftup}. According to this model, the streamwise perturbation amplitude can be expressed as follows: 

\begin{equation}
    u(t) = u_0 \cos{(\beta W_0t)} - v_0U't\label{eq_w0}
\end{equation}

While the classical algebraic growth is retrieved, a new term, due to cross-flow effects, appears. This term induces, for early times, an oscillation of frequency $\beta W_0$ of the amplitude of the streamwise velocity of the perturbation.

\section{Direct numerical simulations}\label{sec:DNS}

The evolution of these two different perturbations have been monitored through DNS yielding two plausible transition scenarios. The simulations have been performed with the spectral element incompressible solver Nek5000 \citep{argonne_national_laboratory_nek5000_2019}. The code is based on a splitting method which leads to solving a set of time dependent problems and a final correction step together with a $P_n-P_{n-2}$ spatial discretisation. Convective terms are treated with an explicit method while linear ones are solved with an implicit scheme. This special type of Robin boundary conditions were implemented following \citet{picella_laminarturbulent_2019}. 

\begin{table}
  \begin{center}
\def~{\hphantom{0}}
  \begin{tabular}{lccccccc}
            & $L_x$ & $L_y$ & $L_z$ & $N_x$ & $N_y$ & $N_z$ & $N_p$ \\[3pt]
        TS case  & 8.9759  & 2 & 10.4720 & 32 & 24 & 26 & 8  \\
        CF case  & 31.4159 & 2 & 3.1415 & 90 & 32 & 10 & 8  
  \end{tabular}
  \caption{Numerical parameters for the two simulations. $N_p$ denotes the polynomial order of the elements.}
  \label{tab:NumParam}
  \end{center}
\end{table}

Two transition scenarii are considered: one initiated with a 3D TS wave having $(\alpha_0, \beta_0) = (0.7,-0.6)$, presented in section \S \ref{sec.3DTS}, and the other with a CF mode having $(\alpha_0, \beta_0) = (0.2,-6)$, discussed in section \S \ref{sec.CF}. This choice of wavenumbers results from a compromise between their distance to the neutral curve and the numerical cost of the simulation, since the computational domain would depend on the perturbation wavenumber. The domain sizes are respectively $[L_x, L_y, L_z] = [2\pi/\alpha_0, 2, 2\pi/\beta_0]$ and $[L_x, L_y, L_z] = [2\pi/\alpha_0, 2, 6\pi/\beta_0]$. In the CF case, a larger computational box in the spanwise direction is used to partially relax the restriction on the wavelengths of developing modes. For the CF scenario, the selected wavenumbers were also chosen not be in the region of important transient growth in an attempt to separate the different instability mechanisms. In both cases, the Reynolds number is $Re=12000$. Both spatial \citep{wassermann_mechanisms_2002, wassermann_transition_2003} and temporal \citep{wintergeste_2000, koch_stability} DNS frameworks are valid and have been considered in studies of swept flows.  All the numerical parameters employed in these two cases can be retrieved in table \ref{tab:NumParam}. Spectral elements in the wall-normal direction have been distributed following a Chebyshev grid to further increase the spatial resolution near the walls.

From a numerical point of view, supercritical transition can be triggered by superposing on the base flow (\ref{BF}) the most unstable wave found from the linear stability analysis, as: 
\begin{equation*}
    \mathbf{U}(\mathbf{x},t=0) = \mathbf{U}_0(\mathbf{x}) + A\mathbf{u}^{}(\mathbf{x})
    \label{BfDns}
\end{equation*} 

where A is the amplitude of the initial perturbation which is set such as, in all the following, $E_c(t=0) \approx 10^{-5}$. This low value for the disturbance energy density ensures a phase of linear exponential growth before nonlinear effects arise.

For further insight, several indicators are monitored through the simulation, most notably the kinetic energy density of the disturbance (as previously defined) and the friction Reynolds number, which is defined as:

\begin{equation}
    Re_{\tau}= \sqrt{Re|\partial_y \overline{U}(\mathbf{x},t)|_{\pm1}}
\end{equation}
where the overbar denotes spatial averaging on the $x-z$ plane. This quantity will brutally increase during the turbulent breakdown while remaining almost constant for both laminar and turbulent flows. In the laminar case, the friction Reynolds number can be found analytically and reads:

\begin{equation*}
    Re_{\tau} = \sqrt{\frac{2Re}{1+3L_{11}}} \approx 150
\end{equation*}

Transition to turbulence can also be investigated through the evolution of the spectral energy associated with Fourier modes \citep{zang_numerical_1989,schmid_stability_2001}: 
\begin{equation}
    \hat{E}_{k_x, k_z} = \frac{1}{2}\int_{-1}^1 |\hat{u}_{k_x,k_z}(y,t)|^2 dV
\end{equation}

where $\hat{u}_{k_x,k_z}(y,t)$ is the Fourier mode of the perturbation velocity field with streamwise and spanwise wavenumbers  $k_x$ and $k_z$, respectively. Following the formalism of \citet{schmid_stability_2001}, the mode denoted as $(m,n)$  corresponds to the $(m \alpha_0, n \beta_0)$ Fourier mode. 
Finally, the transition scenario can also be monitored in the frequency domain through a temporal Fourier transform.

\section{First scenario: 3D TS waves}\label{sec.3DTS}

\subsection{Overview of the transition}

\begin{figure}
    \centering
    \includegraphics[width=0.45\textwidth]{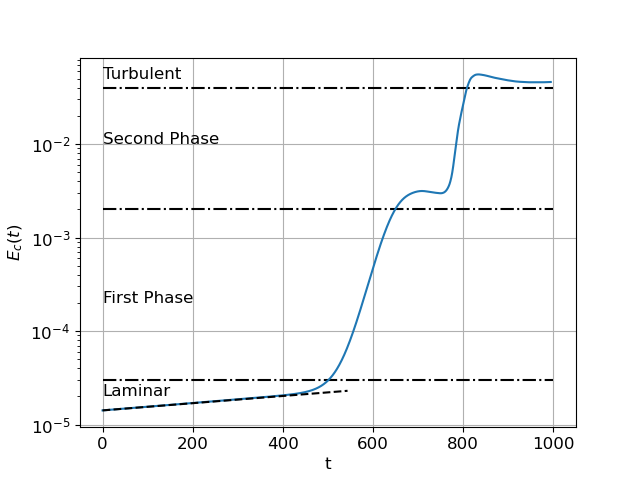}
    \includegraphics[width=0.45\textwidth]{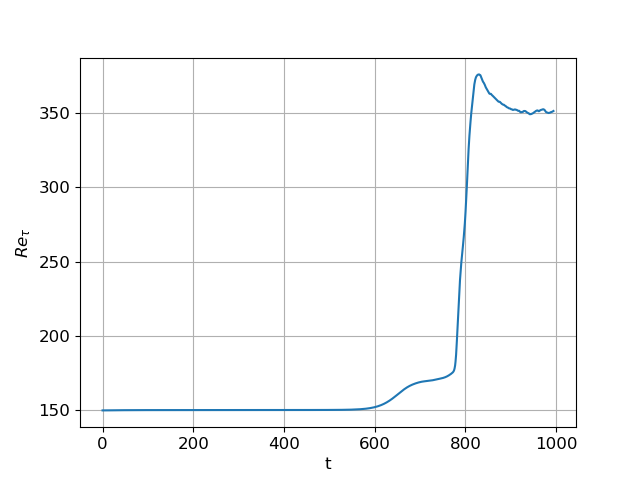}
    \caption{Evolution of the disturbance kinetic energy density (left) and $Re_{\tau}$ (right). The dashed line has a slope $2\omega_r$, where $\omega_r \approx 0.00044$ is the growth rate of the most unstable mode found by means of linear stability analysis.}
    \label{fig:KeReTau}
\end{figure}

\begin{figure}
    \centering
    \includegraphics[scale=0.6]{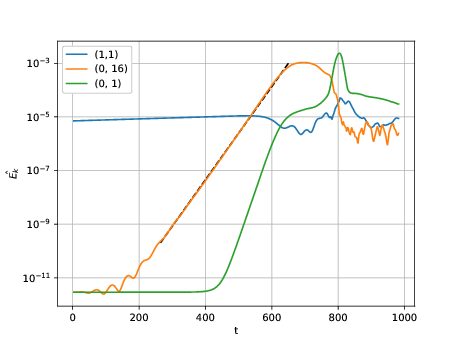}
    \caption{Time evolution of the Fourier modes (1,1), (0,16) and (0,1). The dashed line has a slope $2\sigma \approx 0.04$. Notice the oscillations in the early evolution of mode (0,16).}
    \label{fig:FourierTS}
\end{figure}

\begin{figure}
    \centering
    \includegraphics[width=0.45\textwidth]{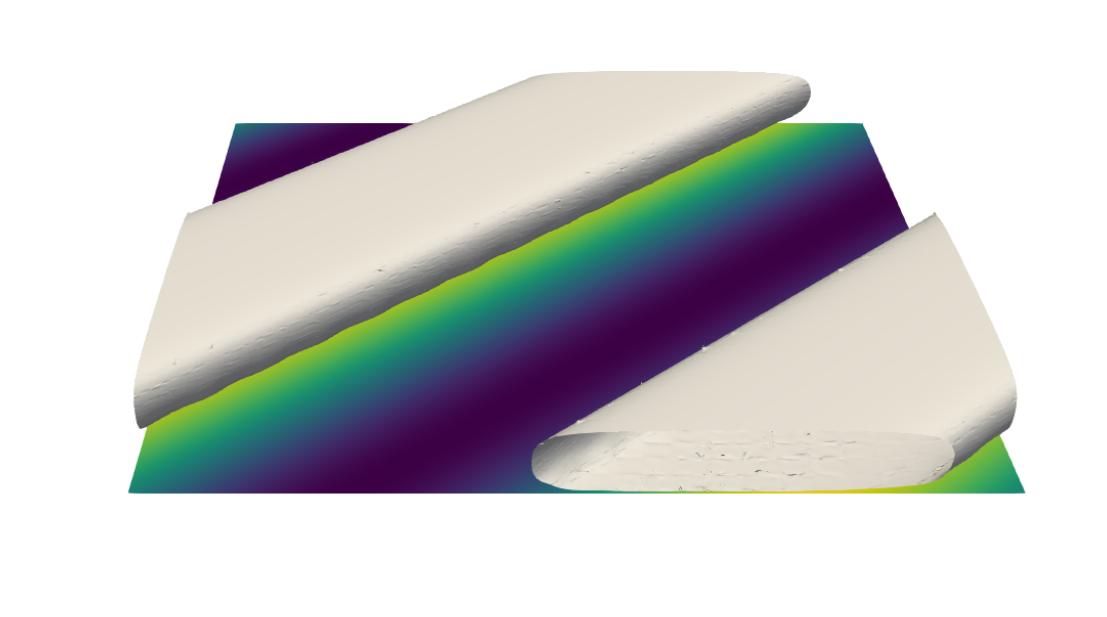}
    \includegraphics[width=0.45\textwidth]{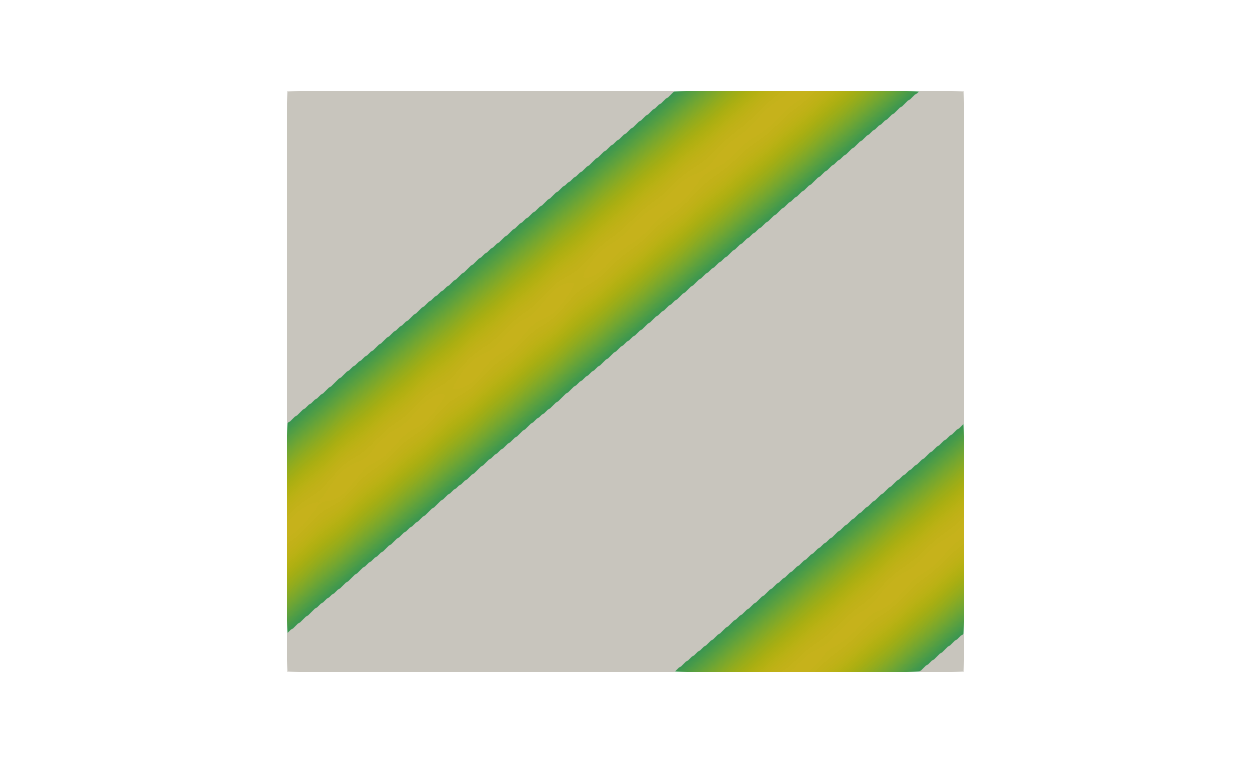} \\
    \includegraphics[width=0.45\textwidth]{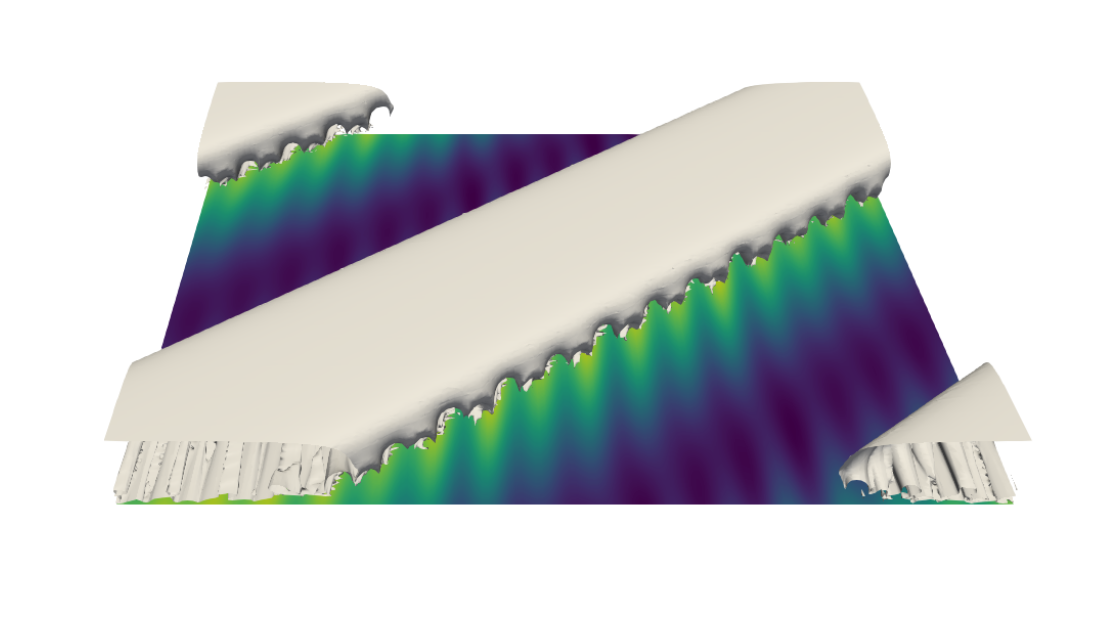}
    \includegraphics[width=0.45\textwidth]{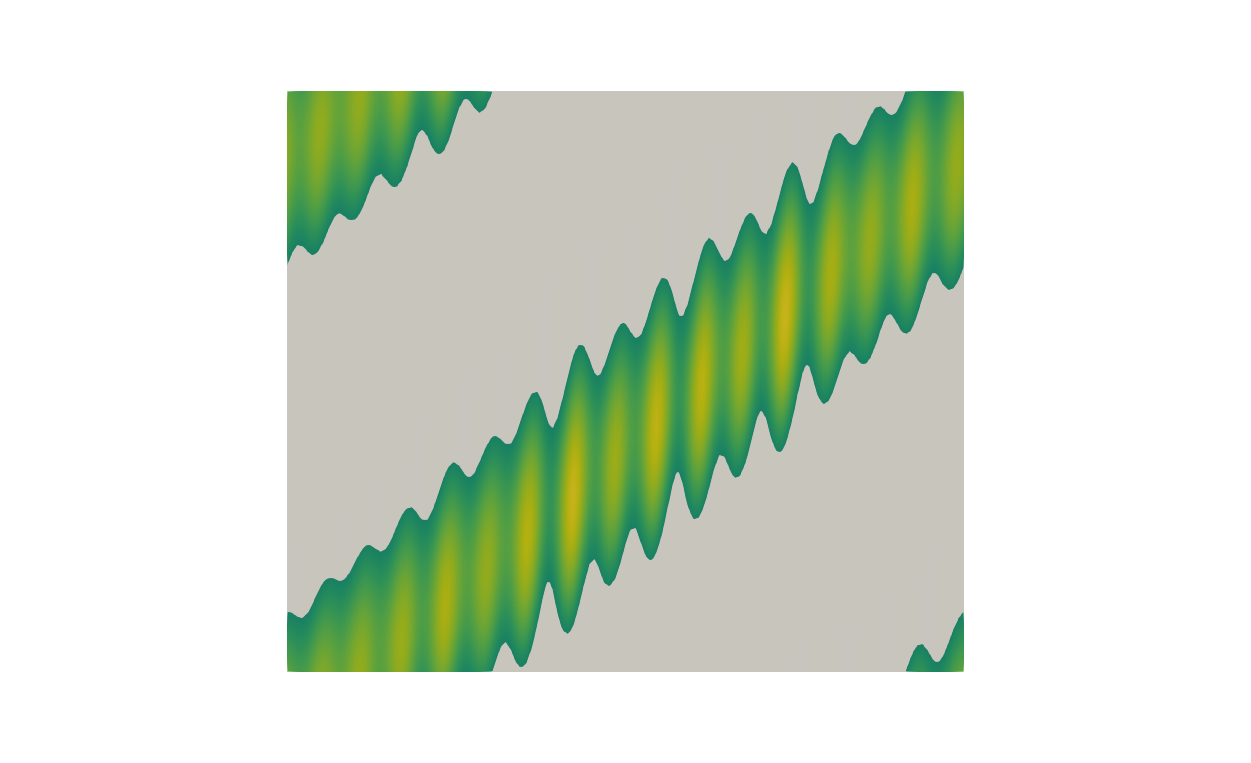} \\
    \includegraphics[width=0.45\textwidth]{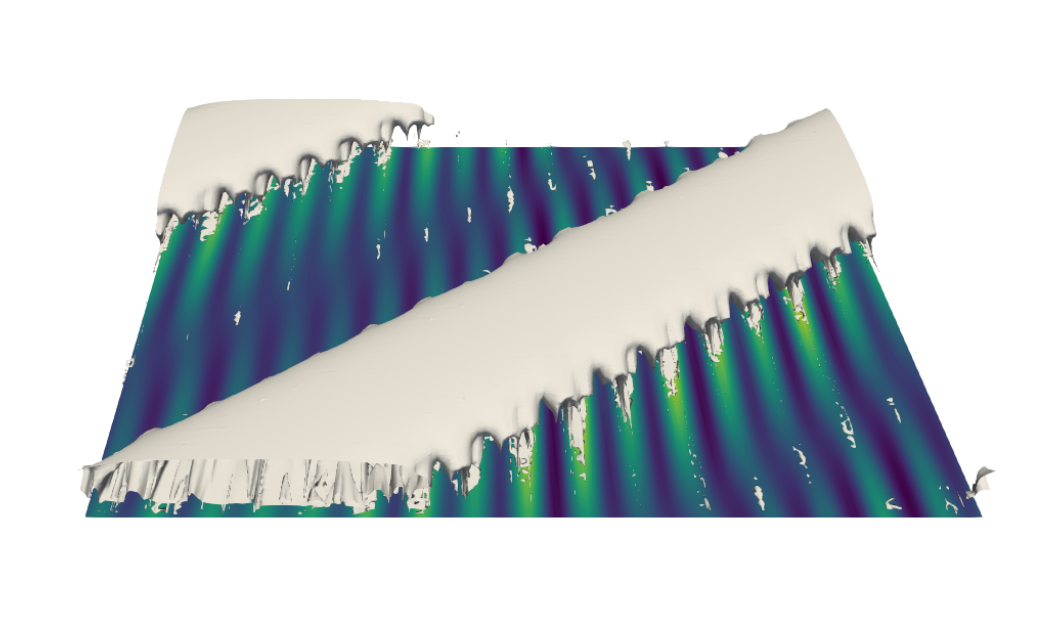}
    \includegraphics[width=0.45\textwidth]{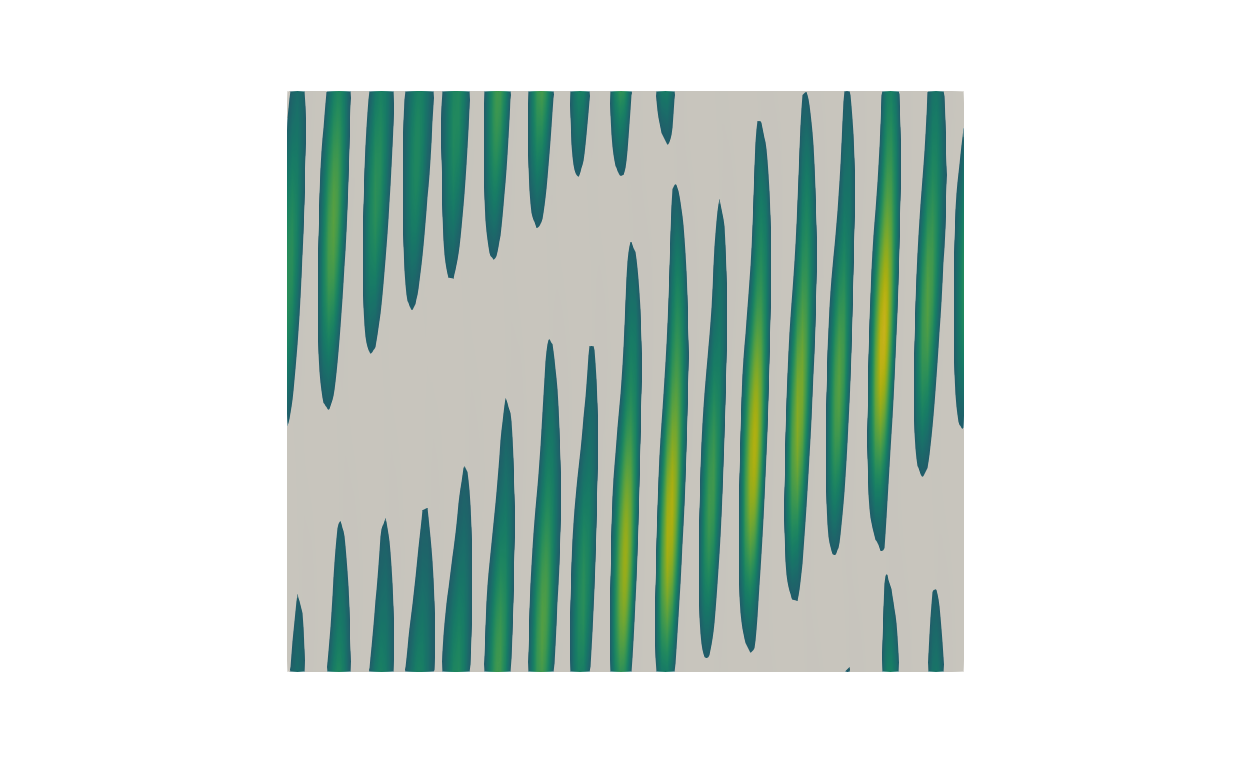} \\
    \includegraphics[width=0.45\textwidth]{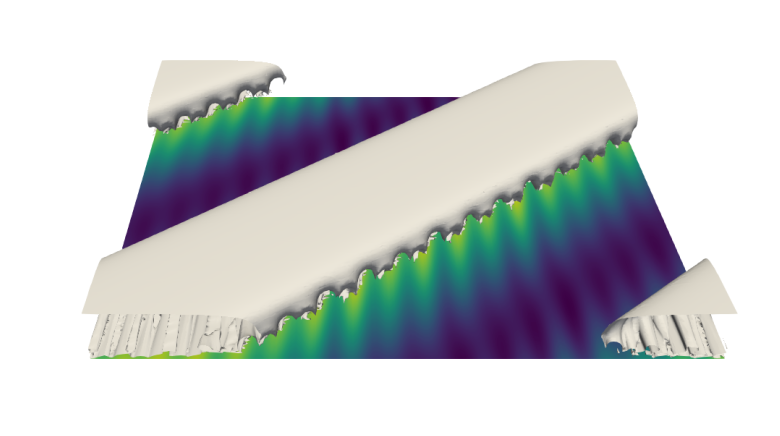}
    \includegraphics[width=0.45\textwidth]{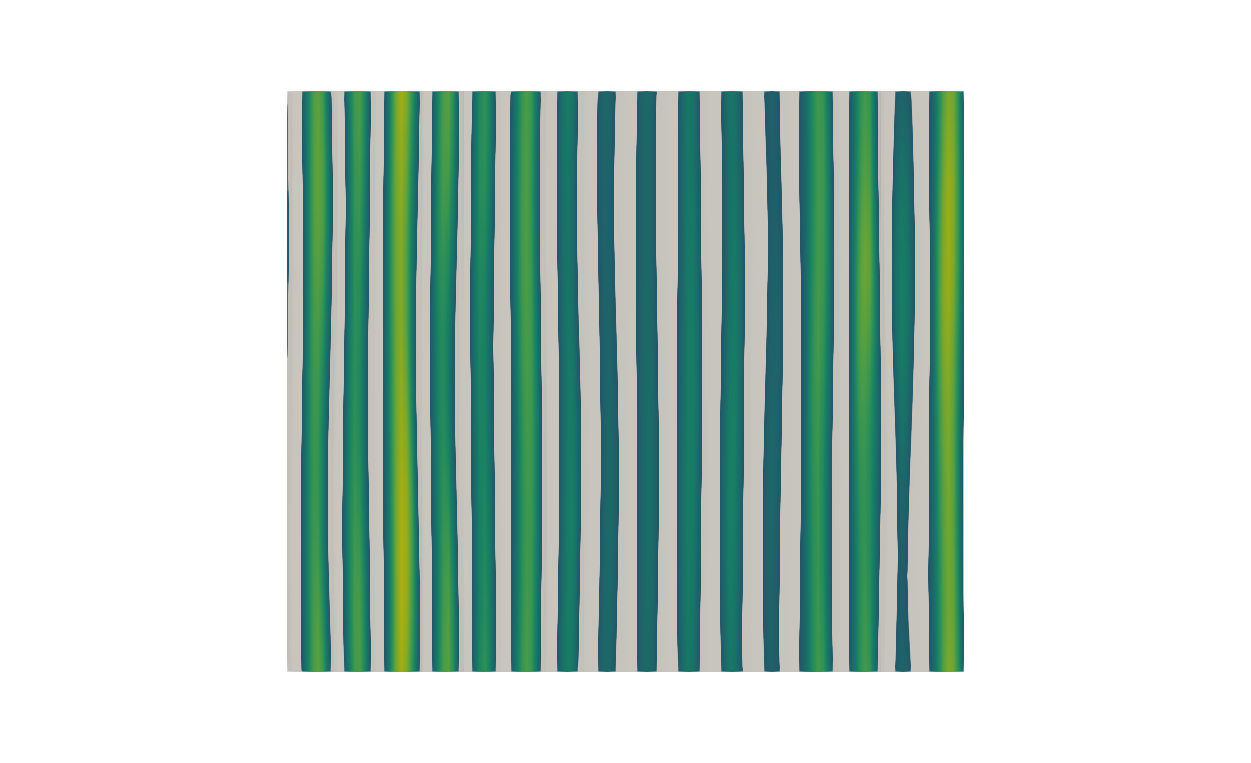} \\
    \includegraphics[width=0.45\textwidth]{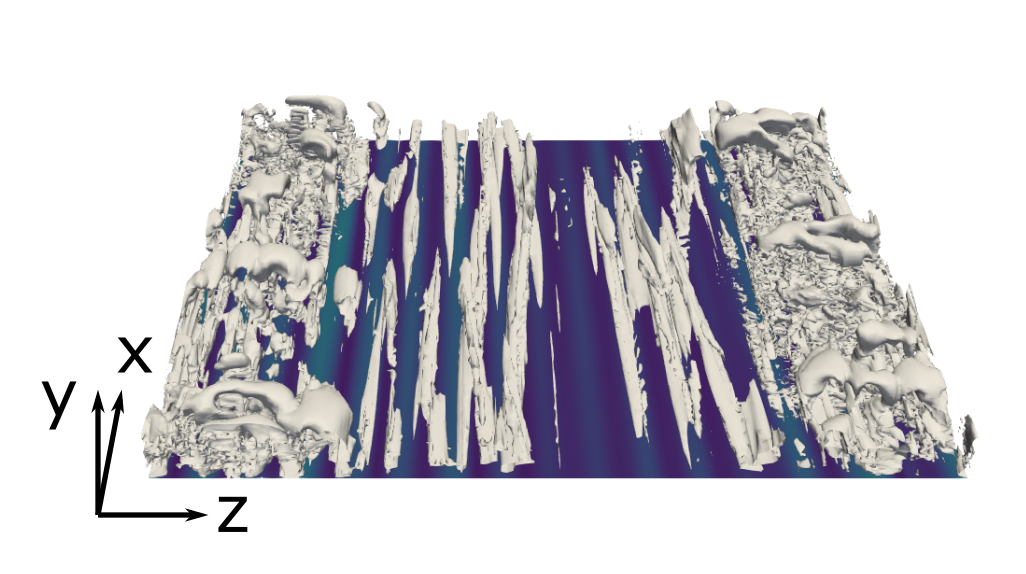}
    \includegraphics[width=0.45\textwidth]{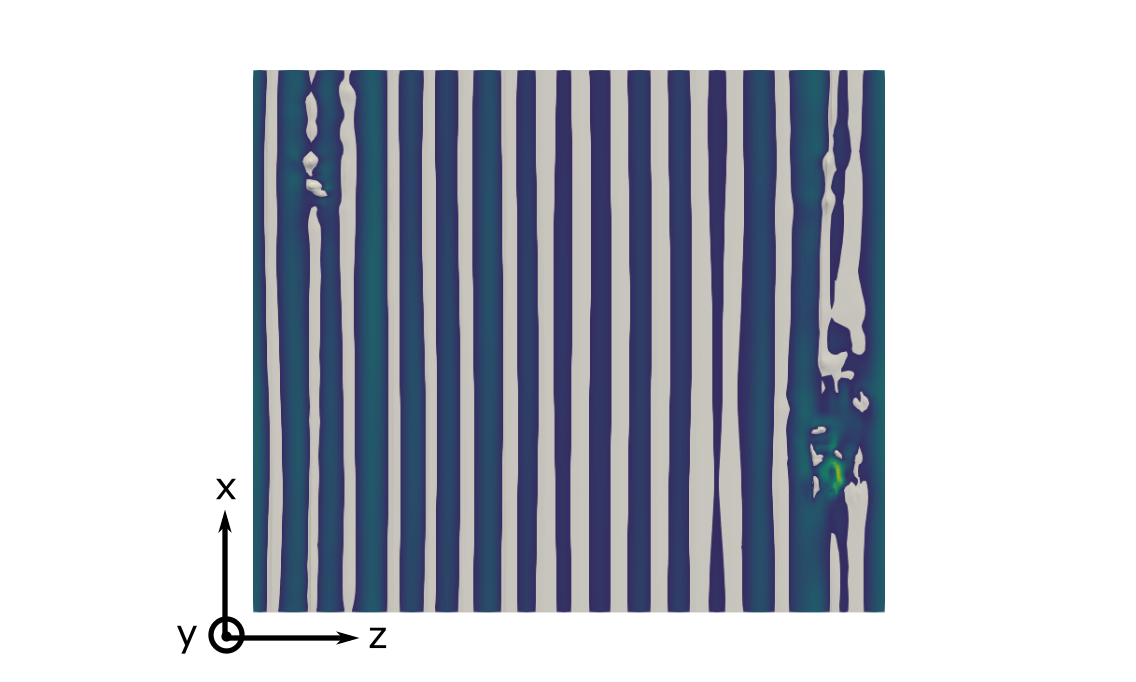} \\
    \caption{Snapshots of the flow at different time T. From top to bottom: $T=200$, $T=400$, $T=600$, $T=800$ and $T=900$. Left: Isosurfaces of the $\lambda_2$-criterion, $\lambda_2=-10^{-5}, -10^{-4}, 0.25$, respectively, and contours of the streamwise velocity at the wall. Right: Isosurface  of the streamwise velocity ($U \approx 0.047$) and contours of $u$ at the wall. The flow is from bottom to top, and left to right. For the sake of clarity, only half channel is shown.}
    \label{fig:SnapshotsTS}
\end{figure}

The transition scenario initiated by a three-dimensional TS wave with $(\alpha_0, \beta_0) = (0.7,-0.6)$ is investigated. The time evolution of the kinetic energy evolution and friction Reynolds number are provided in figure \ref{fig:KeReTau}, while figure \ref{fig:FourierTS} depicts the evolution of the spectral energy for several Fourier spatial modes. In the initial phase, the kinetic energy density displays an exponential increase, with a growth rate in perfect agreement with the linear stability analysis. Figure \ref{fig:FourierTS} shows that the Fourier mode $(1,1)$,  which is the only finite-amplitude mode present in the flow at early times, follows this evolution. Temporal Fourier analysis of this first phase of transition ($0<t<500$), provided in figure \ref{fig:PSDTS}, exhibits a strong peak at $f = 0.024$, corresponding to the frequency of the TS wave. Figure \ref{fig:SnapshotsTS} (top row) shows that at $t=200$, the perturbation is still characterised by a TS wave shape.  

Meanwhile, at $t \approx100$, the mode $(0,16)$ starts oscillating with a period $T_0 = 47$.  The period of the oscillations is in good agreement with that analytically predicted from equation \eqref{eq_w0}, namely $2\pi/(\beta W_0) = 46.58$, hinting at the presence of the modified lift-up previously described. Notice that these oscillations can be observed only on this high wavenumber mode, while other streamwise-invariant Fourier modes such as $(0,1)$ (see figure \ref{fig:FourierTS}) does not display an initial oscillatory phase. A possible reason for this behaviour is that both the target time at which maximum transient growth is recovered, $T_{opt}$, and the oscillation period, $T_0$, are much larger for small spanwise wavenumbers. Respectively, for $(0,16)$, it can be found that $T_{opt} = 100$ and $T_0 = 47$ while $T_{opt} = 615$ and $T_0 = 639$ for the $(0,1)$ mode, which is a too large period to be observed during the linear phase of the perturbation evolution.

After $t\approx280$, the $(0,16)$ mode exhibits exponential growth, characteristic of a secondary instability. At $t\approx520$, the $(0,16)$ mode becomes more energetic than the primary one. At the same time, a brutal increase in the kinetic energy indicates that the first step of transition to turbulence has initiated. This translates into a change in the topology of the mode characterised by the onset of near-wall streamwise-elongated coherent structures, shown in figure \ref{fig:SnapshotsTS} (second and third rows from the top). These streaky structures are characterized by a large spanwise wavenumber, $\beta=-9.6$, corresponding almost exactly to  $16$ times the fundamental spanwise wavenumber $\beta_0$ (namely, $\beta=16\beta_0= -9.6$). Temporal Fourier transform carried out in the time range ($0<t<800$) corresponding to the first phase of transition shows the onset of  low-frequency oscillations at $f=0.007$ (see right frame of figure \ref{fig:PSDTS}) which seems to be associated to such high-wavenumber spatial oscillations.

\begin{figure}
    \centering
    \includegraphics[scale=0.4]{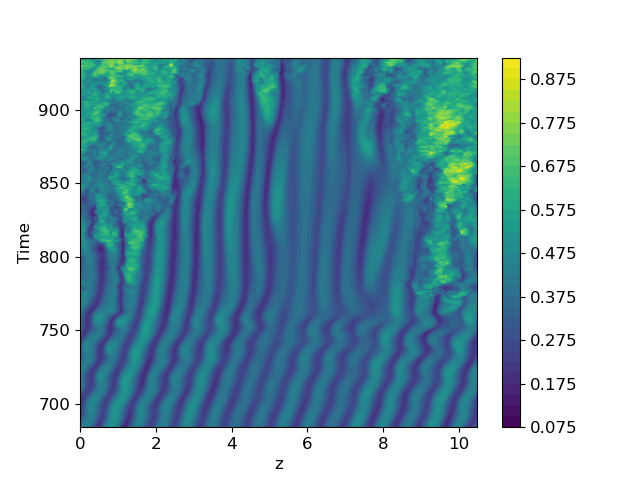}
    \includegraphics[scale=0.4]{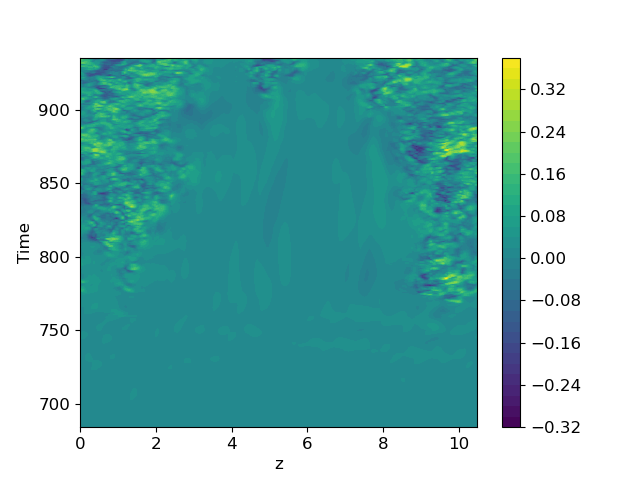}
    \includegraphics[scale=0.4]{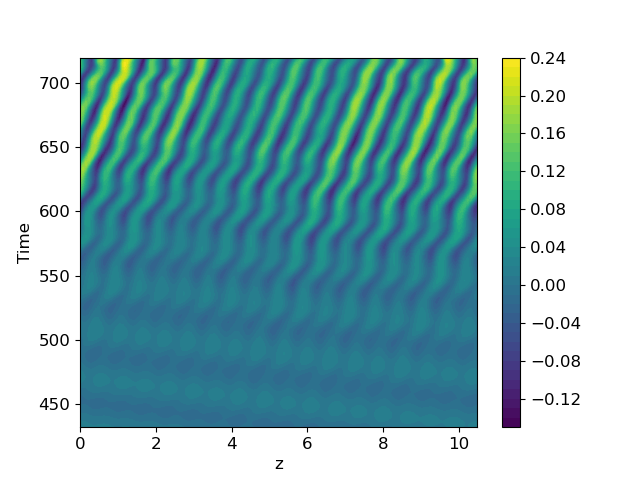}
    \includegraphics[scale=0.4]{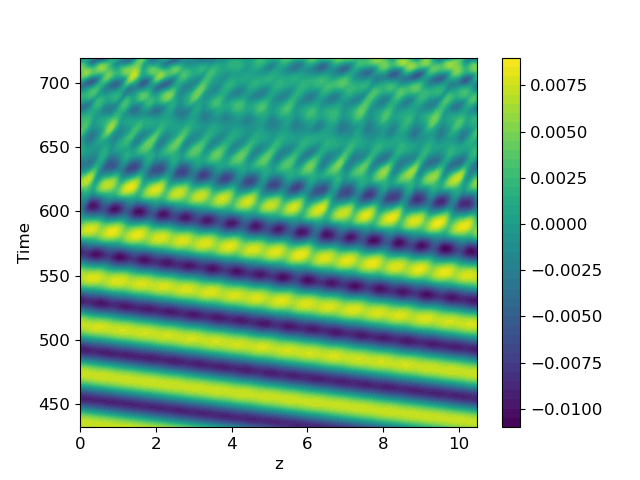}
    \caption{Spatio-temporal evolution of the streamwise (left column) and spanwise (right column) components of the velocity at $x=0$ and $y=-0.85$. For visualisation purposes, the first (second) phase of the transition is depicted separately in the first (second) row. }
    \label{fig:SpatioTemporal}
\end{figure}

\begin{figure}
    \centering
    \includegraphics[scale=0.4]{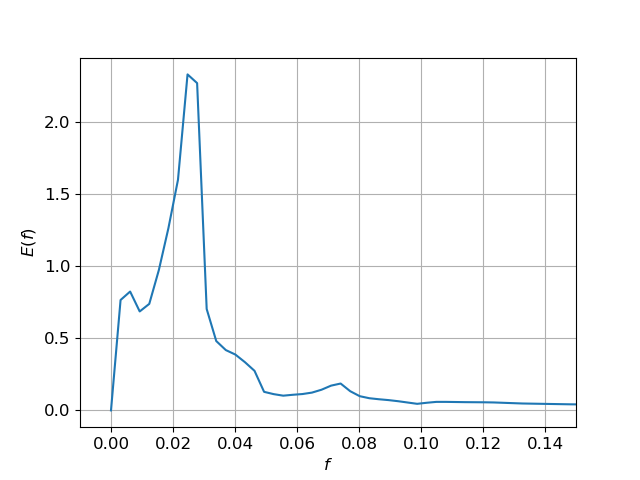}
    \includegraphics[scale=0.4]{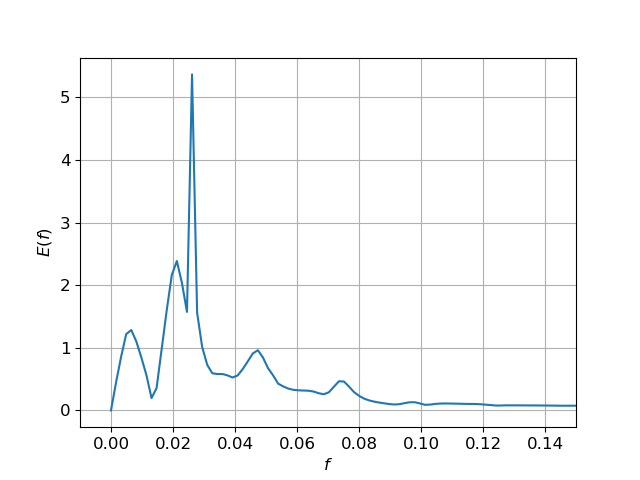}
    \caption{Temporal Fourier energy spectrum of transition to turbulence. The data time-series has been truncated at $t=500$ (left) just after the onset of secondary instability and at $t=800$ (right) before turbulence is established. }
    \label{fig:PSDTS}
\end{figure}

In this transition phase, energy is taken from the 3D TS wave and transferred to these oblique waves, (see figure \ref{fig:SnapshotsTS}, third and fourth rows from the top) ultimately leading to the disappearance of the TS wave. As it can be seen in figure \ref{fig:SpatioTemporal}, showing the spatio-temporal evolution of the streamwise (left) and spanwise (right) components of the velocity,  these streaky structures are not streamwise independent and are oriented with an angle $\phi = 9^o$ with respect to the streamwise direction, almost perpendicular to the initial 3D TS wave. Notice also how the first phase of the transition leads to the suppression of the spanwise velocity component, confirming the streaky nature of the structures. 

At $t = 650$, the kinetic energy reaches a plateau for $E = 3\times10^{-3}$ as the secondary instability saturates (see figure \ref{fig:KeReTau}). Ultimately,  tertiary instability triggers, at $t\approx800$, a dramatic increase in the friction Reynolds number, indicating the breakdown into a fully turbulent state. This final transition can be linked to the $(0,1)$ mode, as suggested by its sudden growth at such time (see figure \ref{fig:FourierTS}). The nature of this final instability is quite complex: it is highly nonlinear as suggested by the presence of harmonics of both primary and secondary modes, which can be observed in the right frame of figure \ref{fig:PSDTS}. It also appears to be, to some extent, related to streak instability. Figure \ref{fig:SpatioTemporal} (bottom row) depicts streaky structures which, quite interestingly, have lost their orientation and are now aligned with the streamwise direction. The streak instability can be observed at time $t\approx700$ in the two bottom rows of figure \ref{fig:SnapshotsTS} and in the bottom left frame of figure \ref{fig:SpatioTemporal}).

Quite notably, this transition scenario is fundamentally different from the usual transition scenario in channel flows. Indeed, supercritical transition to turbulence in channel flow is well documented in the no-slip case \citep{zang_numerical_1989,schmid_stability_2001}: after an initial exponential growth phase, secondary instability of a subharmonic mode appears. The interaction between the TS wave and the secondary modes leads to a peak-valley structure \citep{herbert_secondary_1985,asai_origin_1989} which, in turn, results in the formation of a staggered pattern of $\lambda$-vortices. Then, $\lambda$-vortices subsequently develop into hairpin vortices that will ultimately breakdown into turbulence. In the slip case, with a homogeneous Navier slip boundary condition, \cite{picella_laminarturbulent_2019} have shown that a delay of the laminar-turbulent transition is possible, but the nature of the instability mechanisms remains virtually the same.

In the present case, the cross-flow component of the base flow appears to have a dramatic effect on the transition scenario. Similarly to what we have reported concerning the primary instability, this cross-flow component might originate a new secondary instability mechanism, which would be investigated in the next subsection. If these cross-flow related secondary modes have higher growth rates than the subharmonic ones leading to $\lambda_2$ vortices \citep{herbert_secondary_1985}, the secondary phase of transition would be dominated by cross-flow related mechanisms. 

As observed in swept flows \citep{serpieri_cross-flow_2018}, during this secondary phase, oblique waves appear and eventually, through nonlinearities, saturate. The saturated cross-flow vortices contain several strong shear layers, which can, under certain circumstances, further destabilise. The most common instability mechanism observed in such flows \citep{bippes_instability_1990, bippes_basic_1999} involves the shear layer at the bottom of the vortex, created from the circulation of high-speed fluid toward lower velocities region near the wall.

In the present case, the slip boundary conditions reduce the efficiency of this mechanism by lowering the wall normal velocity gradients near the walls. Instead, the flow keeps accelerating in the spanwise direction until it reaches a neighbouring vortex, thus creating stagnation points and strong shear layers in the spanwise direction. Transient growth in this region leads to the formation of streaks \citep{guegan_huerre_schmid_2007} aligned with the streamwise direction. Streak instability concludes the transition to turbulence.

This scenario of transition will be corroborated in the next subsection, by means of secondary stability analysis and comparison with DNS.

\subsection{Floquet stability analysis}

In an effort to provide a more qualitative description of the secondary phase of the transition, a secondary stability analysis, based on Floquet theory, is realised. This method has been applied to both channel flows \citep{herbert_secondary_1983} and two-dimensional boundary layers \citep{herbert_secondary_1985} for which it has successfully predicted the secondary instability of a 2D TS wave. An extension to three dimensional base flows was proposed and applied most notably to swept wings \citep{fischer_primary_1991,janke_secondary_2000, liu_floquet_2008}. Secondary stability equations in a velocity-vorticity formulation can be found in \citet{fischer_primary_1991}. Whereas, a primitive variables formulation for 2D base flows only, can be found in \citet{schmid_stability_2001}. 

In this work, due to the three-dimensionality of the problem under consideration, we use a formulation of Floquet theory and subsequent modal expansion somewhat different (but nevertheless equivalent) from those used in these articles. In \cite{herbert_secondary_1983} study, Squire theorem ensures a 2D initial perturbation while, for swept flows cases, an initial 3D perturbation should be considered. For this reason,  Floquet theory for swept flows is usually applied in a rotated frame aligned with the 3D wave direction of propagation, thus effectively reducing the base flow to two dimensions. Precisely, the base flow for secondary stability analysis is constructed as a superposition of an unstable mode and the laminar flow profile such as: 

\begin{equation}
    \mathbf{U}_1(x,y,t) = \mathbf{U}_0(y) + A\mathbf{\hat{u}}^{TS}(y)\e^{i\alpha_0x + i\beta_0 z -i\omega_0t}
    \label{Floquet_BF}
\end{equation}

with $A$ the amplitude of the unstable 3D TS wave obtained from linear stability analysis. Considering that the TS wave moves at the velocity $c_x$ and $c_z$ in the $x'$ and $z'$ directions, respectively, the time dependence of the base flow is removed when the moving frame $(x',z') = (x-c_xt, z-c_zt)$ is considered. Linearising the Navier-Stokes equations around this new base flow in the co-moving frame leads to a system of ODEs (see Appendix \ref{app:secondary}) with $2\pi/\alpha_0$-periodic and $2\pi/\beta_0$-periodic coefficients in the $x$- and $z$-directions, respectively. Thus, Floquet theorem is applied in both the streamwise and spanwise directions:

\begin{equation}
    \mathbf{q_1}(x',y,z',t) = \mathbf{\Tilde{q}}(x',y,z') \e^{\gamma x'} \e^{\mu z'} \e^{\sigma t}
\end{equation}

with $\gamma = \gamma_r + i\gamma_i$ and $\mu = \mu_r + i\mu_i$ the Floquet parameters in the streamwise and spanwise directions. Then, $\mathbf{\Tilde{q}}(x',y,z')$ is Fourier transformed in both the $x'$ and $z'$-directions and introducing the detuning factors $\epsilon = \sigma_i/\alpha_0$, $\delta = \mu_i/\beta_0$, the general form of the solution can be expressed as:

\begin{equation}
    \mathbf{q_1}(x',y,z',t) = \e^{\sigma t} \e^{\gamma _r x'} \e^{\mu_r z'} \sum_{m,n=-\infty}^{+\infty} \mathbf{\Tilde{q}}_{m,n}(y) \e^{i\alpha_0(m + \epsilon)x' + i\beta_0(n + \delta)z'}
    \label{Floquet}
\end{equation}

\begin{figure}
    \centering
    \includegraphics[width=0.45\textwidth]{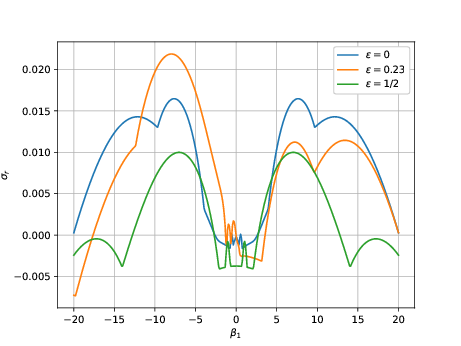}
    \includegraphics[width=0.41\textwidth]{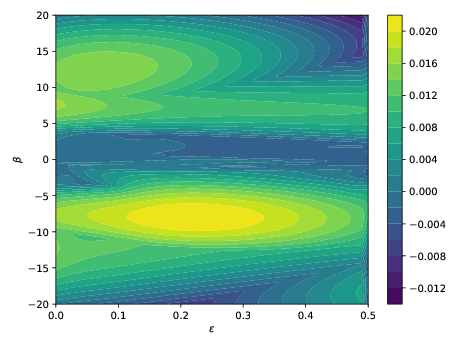}
    \caption{Left: Growth rate $\sigma_r$ of the most unstable mode as a function of the spanwise wavenumber $\beta_1$ for $\epsilon =0$ (fundamental), $\epsilon = 0.22$ (detuned) and $\epsilon = 1/2$ (subharmonic). Right: Contour plot of the growth rate $\sigma_r$ of the most unstable mode for $\beta_1 = -9.6$  as a function of both the detuning factor $\epsilon$ and the spanwise wavenumber $\beta_1$.}
    \label{fig:GrowthSecond}
\end{figure}

When a numerical solution is sought, the modal expansion \eqref{Floquet} needs to be truncated, usually with the lowest possible number of modes. Unfortunately, reaching the spanwise wavenumbers observed in the DNS would require a large number of modes in the $z'$-direction, which make the problem too computationally expensive. In order to relax this constraint, the modal expansion in the spanwise direction is reduced to only one mode by fixing $n$ and $\delta$ and introducing $\beta_1 = \beta_0(n + \delta)$, which represents the effective secondary spanwise wavenumber. Since $n \in \mathbb{Z}$ and $-1/2 < \delta \le 1/2$, the quantity $n+\delta$ spans all real numbers bijectively such that every spanwise wavelength is accessible through a unique choice of $n$ and $\delta$.

Temporal stability is investigated, thus $\gamma_r=\mu_r=0$ and the real part of $\sigma$ indicates the growth rate of the secondary instability. Finally, the modal decomposition \eqref{Floquet} reduces to:

\begin{equation}
    \mathbf{q_1}(x',y,z',t) = \e^{\sigma t} \e^{i\beta_1 z'} \sum_{m=-\infty}^{+\infty} \mathbf{\Tilde{q}}_{m}(y) \e^{i\alpha_0(m + \epsilon)x'}
    \label{Floquet2}
\end{equation}

Introducing \eqref{Floquet2} into the linearised Navier-Stokes equations leads to an infinite set of equations which, once truncated, can be recast in an eigenvalue problem likewise the primary stability problem. 

All the results given here are obtained with the lowest possible truncation that is $m=0,1$ in the subharmonic ($\epsilon=1/2$) cases and $m=-1,0,1$ for the fundamental ($\epsilon = 0$) and detuned ($0 < \epsilon < 1/2$) modes. For the sake of clarity, the derivation of the secondary stability equations and the eigenvalue problem is fully detailed in Appendix \ref{app:secondary}.

In both classic channel and boundary layer flows, the detuning factor is not relevant \citep{herbert_secondary_1983, herbert_secondary_1985}: subharmonic instabilities have stronger growth rates than both fundamental and detuned modes. However, this might not be true in our framework, due to the strong asymmetry present in the base flow.

The dependence of the growth rate of the most unstable mode on the detuning factor is investigated in Figure \ref{fig:GrowthSecond}, for $Re=12000$. A large secondary secondary growth rate can be reached for detuned modes and large spanwise wavenumbers, in the range of those recovered for the primary CF instability. The maximum secondary growth rate is $\sigma_r = 0.023$ for $\epsilon = 0.23$ and $\beta_1=-8$. but the instability region is quite large and similar growth rates can be reached for larger $\beta_1$. Precisely, the growth rate corresponding to the spanwise wavelength observed in the DNS (namely, $\beta = -9.6$), is equal to $\sigma_r = 0.020$.

\begin{figure}
    \centering
    \includegraphics[width=0.6\textwidth]{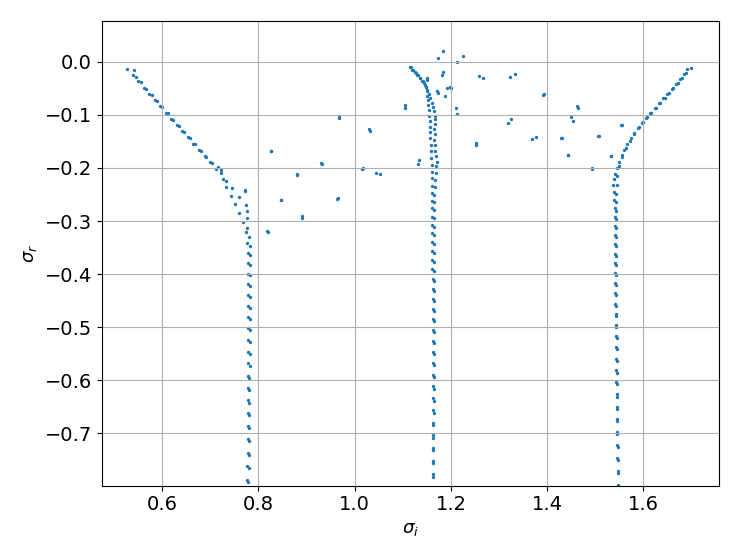}
    \caption{Secondary instability spectrum for $Re=12000$, $\beta_1=-9.6$, $A = 10^{-5}$ and $\epsilon=0.23$. The values of the frequency and growth rate of the four most unstable modes are given in table \ref{tab:ModesSecond}.}
    \label{fig:SpectrumSecond}
\end{figure}

\begin{table}
  \begin{center}
\def~{\hphantom{0}}
  \begin{tabular}{lccc}
            & $\sigma_r$ & $\sigma_i^{TS}$ & $\sigma_i^{0}$ \\[3pt]
        Mode 1  & 0.0205 & 1.1835 & 0.0449  \\
        Mode 2  & 0.0115 & 1.2245 & 0.0849  \\
        Mode 3  & 0.0070 & 1.1729 & 0.0349  \\
        Mode 4  & 0.0004 & 1.2118 & 0.0749
  \end{tabular}
  \caption{Most unstable eigenvalues of the secondary stability analysis for $Re=12000$, $\beta_1 = -9.6$, $\epsilon = 0.23$ and $A = 10^{-5}$. For a single mode, two frequencies, $\sigma_i^{TS}$ and $\sigma_i^{0}$, are given: the former is the frequency of the mode in the frame moving with the TS wave while the latter corresponds to the frequency in the laboratory frame.}
  \label{tab:ModesSecond}
  \end{center}
\end{table}

The spectrum obtained for $Re=12000$, $\beta_1 = -9.6$, $\epsilon = 0.23$ and $A = 10^{-5}$, shown in figure \ref{fig:SpectrumSecond}, exhibits three unstable and one marginally unstable mode. All the modes share a similar structure: the $m=0$ mode, yielding an effective streamwise wavenumber $\alpha = \epsilon\alpha_0 = 0.16$, is dominant in comparison with the $m=-1$ and $m=1$ modes. The wavenumber in the streamwise direction is too small to be retrieved in the computational domain thus explaining, to some extent, the onset of streaky structures. The eigenvalues associated to these modes are summarised in table \ref{tab:ModesSecond}. Considering the dominant $m=0$ mode, frequencies in the laboratory frame are retrieved through the following equation:
\begin{equation}
    \sigma_i^{0} = \sigma_i^{TS} - \beta_1 c_z - \alpha_0 \epsilon c_x.
\end{equation}

\begin{figure}
    \centering
    \includegraphics[width=0.45\textwidth]{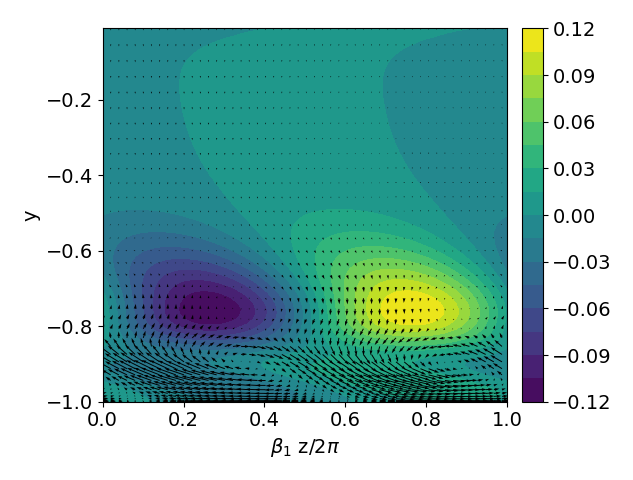}
    \includegraphics[width=0.45\textwidth]{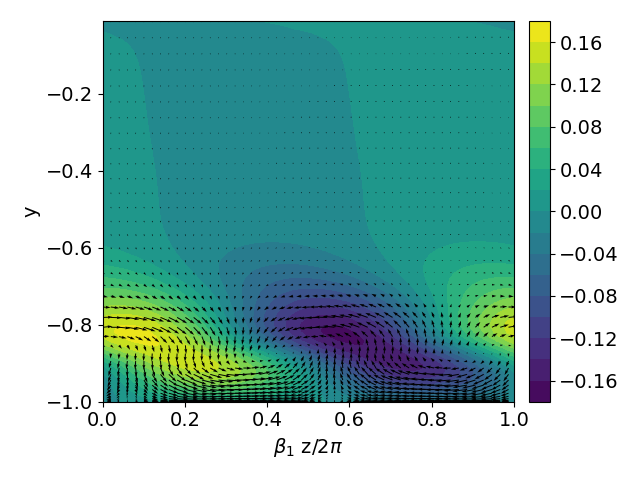} 
    \includegraphics[width=0.45\textwidth]{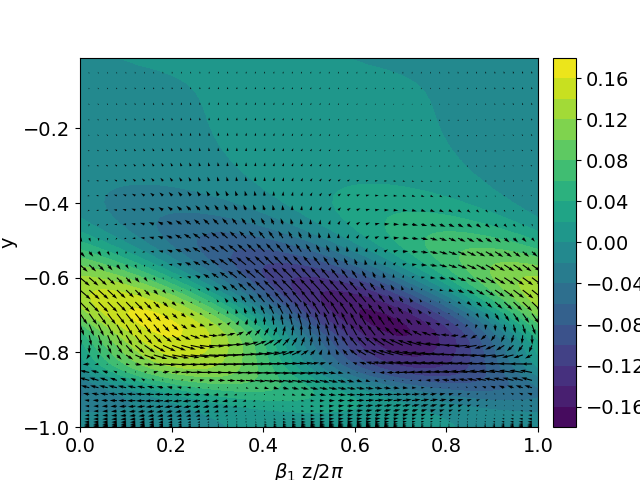}
    \includegraphics[width=0.45\textwidth]{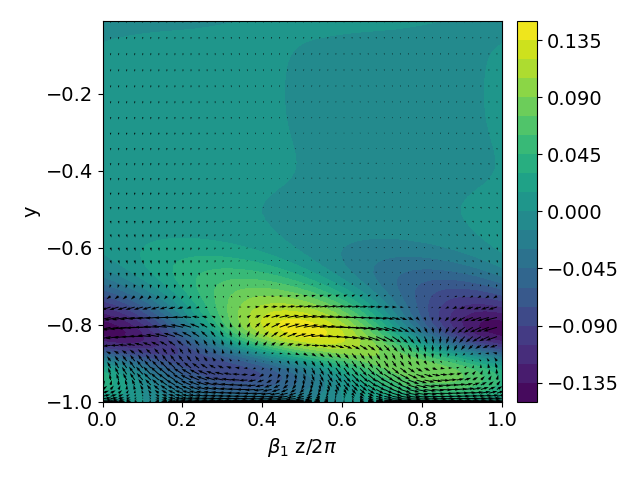}
    \caption{Slice at $x=L_x/2$ of the reconstructed secondary perturbations of the four most unstable modes for $Re=12000$, $\beta_1 = -9.6$, $\epsilon = 0.23$ and $A = 10^{-5}$. From top to bottom and left to right, modes are ordered from the most unstable to the least unstable. The contour plot represents the streamwise velocity disturbance while the quiver plot shows the $v-w$ cross-flow.}
    \label{fig:SecondModes}
\end{figure}

For the four most unstable modes, secondary perturbations are reconstructed based on equation \eqref{Floquet}. A slice of these modes in the $y-z$ plane is provided in figure \ref{fig:SecondModes}, showing that they share a rather similar structure. In the spanwise direction, one can see alternated high and low streamwise-velocity patches. The streamwise velocity component is one order of magnitude higher than the cross-flow components, and is strongly localised in the region where the amplitude of the TS wave reaches a maximum. Cross-flow components form vortical structures concentrated near the walls and similar to the ones found for cross-flow instabilities. More precisely, Mode 1 consists in oblique vortices together with oblique streaky structures for the streamwise velocity component. Modes 2 and 4 are almost identical besides a phase shift. One can also observe a stagnation point near the wall at $z=0.58$ for Mode 2. Mode 3 appears peculiar as its vortices sit on top of a region of low velocity. For all these modes, the vortices push high-momentum fluid towards the wall where, due to the slip boundary conditions, it is strongly accelerated in the spanwise direction. When this high-momentum flow reaches the neighbouring vortices, it is re-ejected upwards back in the flow, creating stagnation points and strong spanwise shear layers in the process.

\begin{table}
  \begin{center}
\def~{\hphantom{0}}
  \begin{tabular}{lccc}
                 & $\beta_1$ & $\sigma_r$   &   $\sigma_i^o$ \\[3pt]
        DNS      &  -9.6  & 0.02  & 0.03958  \\
        Floquet  &  [-6,-10]  & 0.02 & 0.0449
  \end{tabular}
  \caption{Comparison between wavenumbers and growth rates obtained from the DNS via Fourier transform and from secondary stability analysis. }
  \label{tab:Floquet}
  \end{center}
\end{table}

The main features of the most unstable modes recovered by Floquet analysis are compared with those observed in the DNS in table \ref{tab:Floquet}. For the DNS, the secondary growth rate is extracted from figure \ref{fig:FourierTS} while the main temporal frequency is obtained from figure \ref{fig:PSDTS}. As previously discussed, secondary stability analysis yields a range of unstable spanwise wavenumbers  which includes the main wavenumber observed in the DNS. The growth rates are in perfect agreement. The small discrepancy found between the frequencies could arise from the fact that the shape assumption is likely not completely valid and the mean flow is slightly distorted during the transition. Also, only the frequency of the dominant mode $m=0$ was considered while, formally, each mode of the Floquet expansion has its own frequency in the laboratory frame.

\begin{figure}
    \centering
    \includegraphics[width=0.47\textwidth, height=0.5\textwidth]{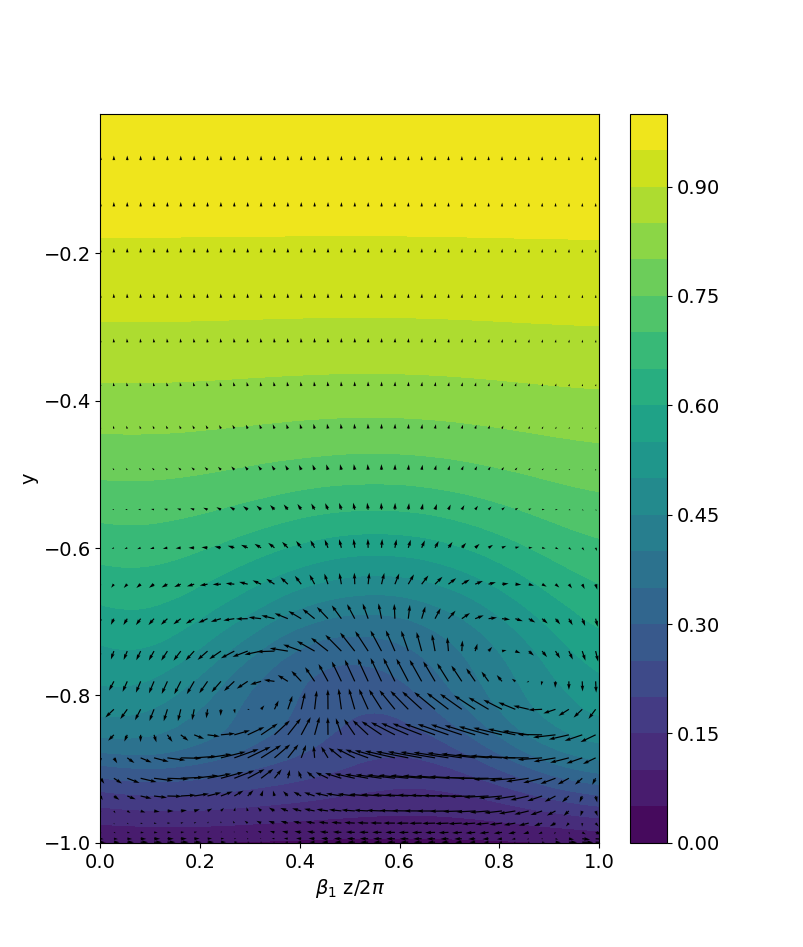}
    \includegraphics[width=0.42\textwidth, height=0.5\textwidth]{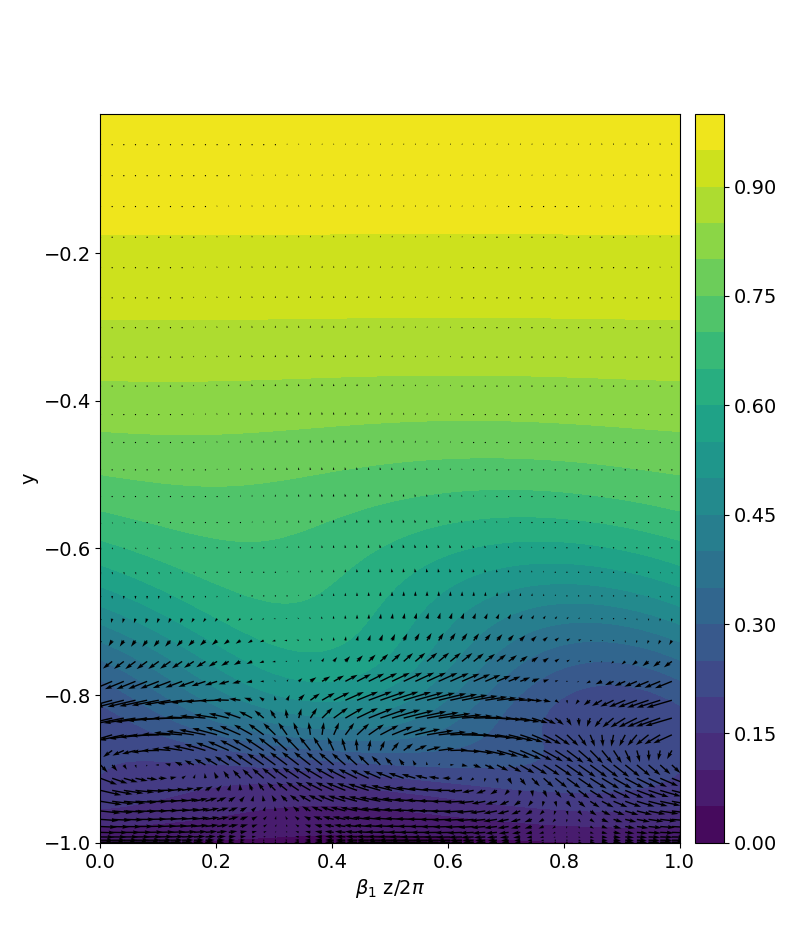}
    \includegraphics[width=0.47\textwidth, height=0.5\textwidth]{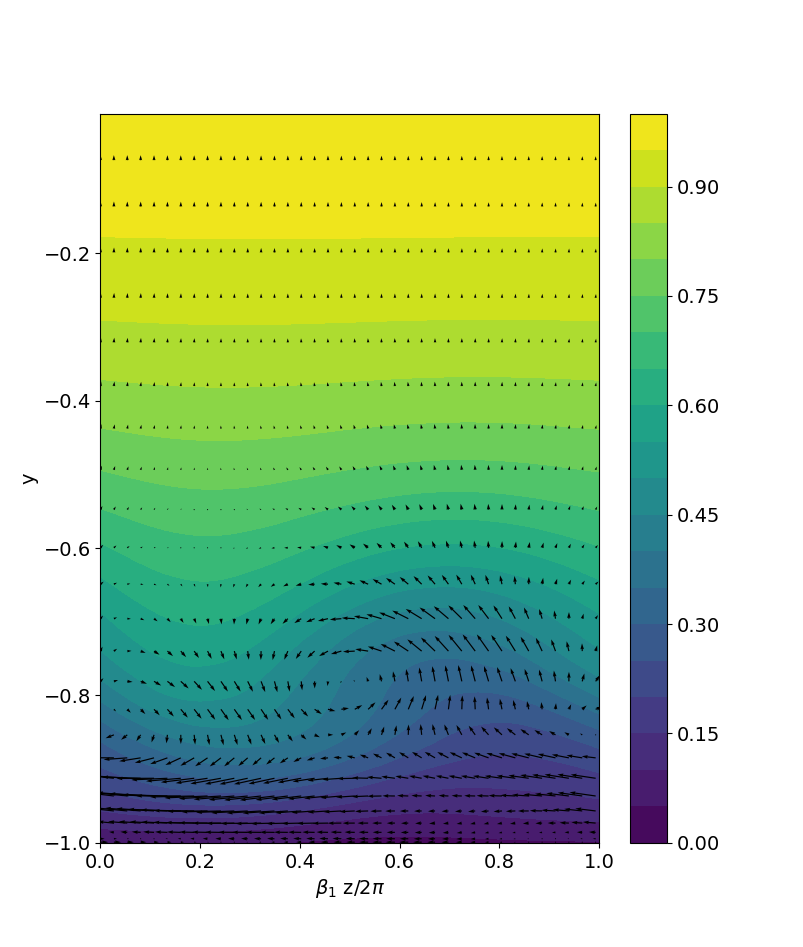}
    \includegraphics[width=0.42\textwidth,height=0.5\textwidth]{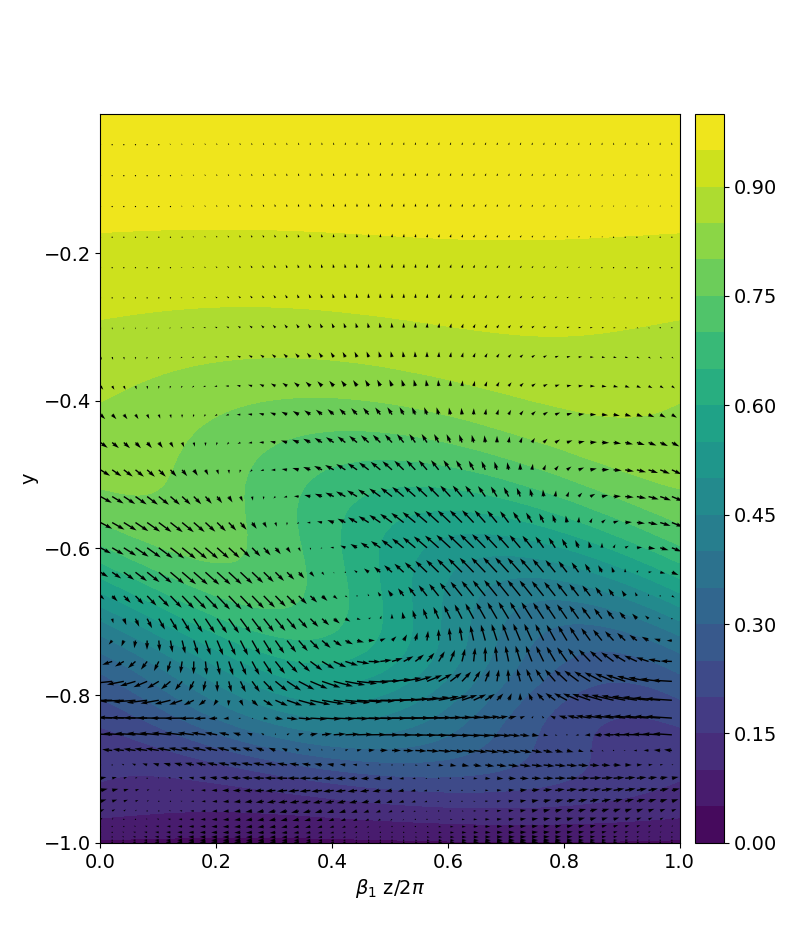}
    \caption{Comparison between snapshots extracted from the DNS (left column) and secondary stability modes (right column): Mode 1 (top row) and Mode 3 (bottom row). Secondary perturbations from the DNS are extracted from snapshots taken at $t=500$, by subtracting the base flow and the primary disturbance. The contour plot represents the total streamwise velocity disturbance while the quiver plot shows the $v-w$ cross-flow. }
    \label{fig:ComparisonDNS}
\end{figure}

Finally, a comparison of the spatial structure of the secondary perturbations obtained from the DNS and some of the modes obtained by secondary stability analysis is provided in figure \ref{fig:ComparisonDNS}. Snapshots of the flow are taken at at $x=L_x/2$ and $t=500$. The latter value is chosen for ensuring that nonlinearities have a negligible effect, while the position in the streamwise direction is the same of that used for the secondary stability analysis. Both the base flow and the initial TS wave are subtracted from the snapshot to retrieve the secondary perturbations. The snapshot is then divided into portions of size $2\pi/\beta_1$ to isolate vortical structures. The spatial structures compare rather well, although for Mode 1 (top row), one can observe that in the DNS the counter rotating vortices are located rather farther from the wall. Regarding Mode 3, its location in the wall-normal direction can be retrieved despite the size of the vortex is slightly smaller. 

\section{Second scenario: cross-flow modes}\label{sec.CF}

\subsection{Overview of the transition}

\begin{figure}
    \centering
    \includegraphics[width=0.45\textwidth]{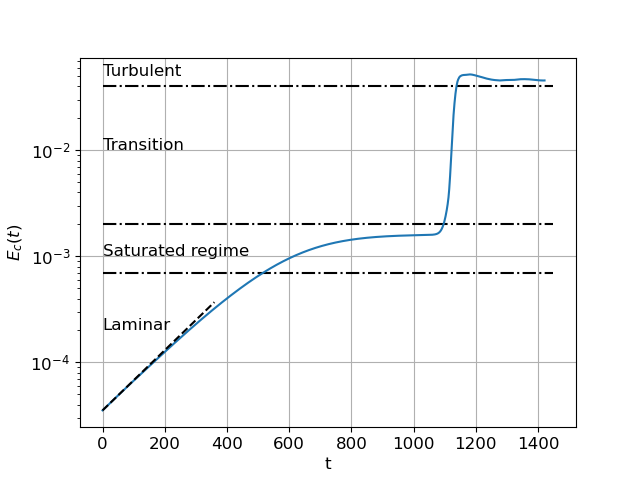}
    \includegraphics[width=0.45\textwidth]{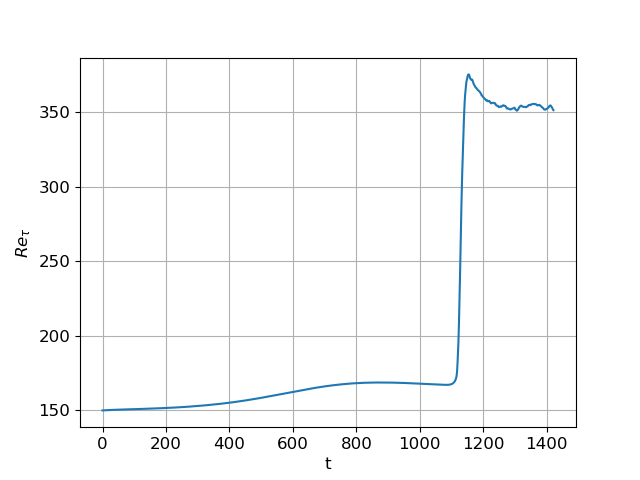}
    \caption{Evolution of the disturbance kinetic energy density (left) and $Re_{\tau}$ (right) for the CF transition scenario. The dashed line has a slope $2\omega_r$ where $\omega_r \approx 0.003267$ is the growth rate of the most unstable perturbation found with the linear stability analysis.}
    \label{fig:KeReTauCF}
\end{figure}
\begin{figure}
    \centering
    \includegraphics[width=0.45\textwidth]{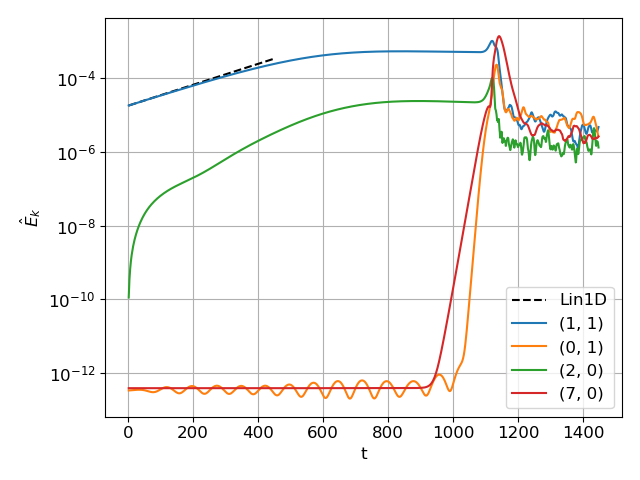}
    \caption{Time evolution of the Fourier modes $(1,1)$, $(2,0)$, $(7,0)$ and $(0,1)$ for the CF transition scenario. The dashed line corresponds to the exponential primary growth rate.}
    \label{fig:FourierCF}
\end{figure}

The second transition scenario is initiated injecting on the base flow the cross-flow unstable mode with $(\alpha_0, \beta_0) = (0.7,-0.6)$. 
%
The time evolution of the kinetic energy evolution and friction Reynolds number is provided in figure \ref{fig:KeReTauCF}, while figure \ref{fig:FourierCF} depicts the evolution of the spectral energy for several Fourier spatial modes. 

\begin{figure}
    \vspace*{-1.7cm}
    \centering
    \includegraphics[trim={3cm 1cm 2cm 2cm},clip,width=0.65\textwidth,height=0.2\textheight]{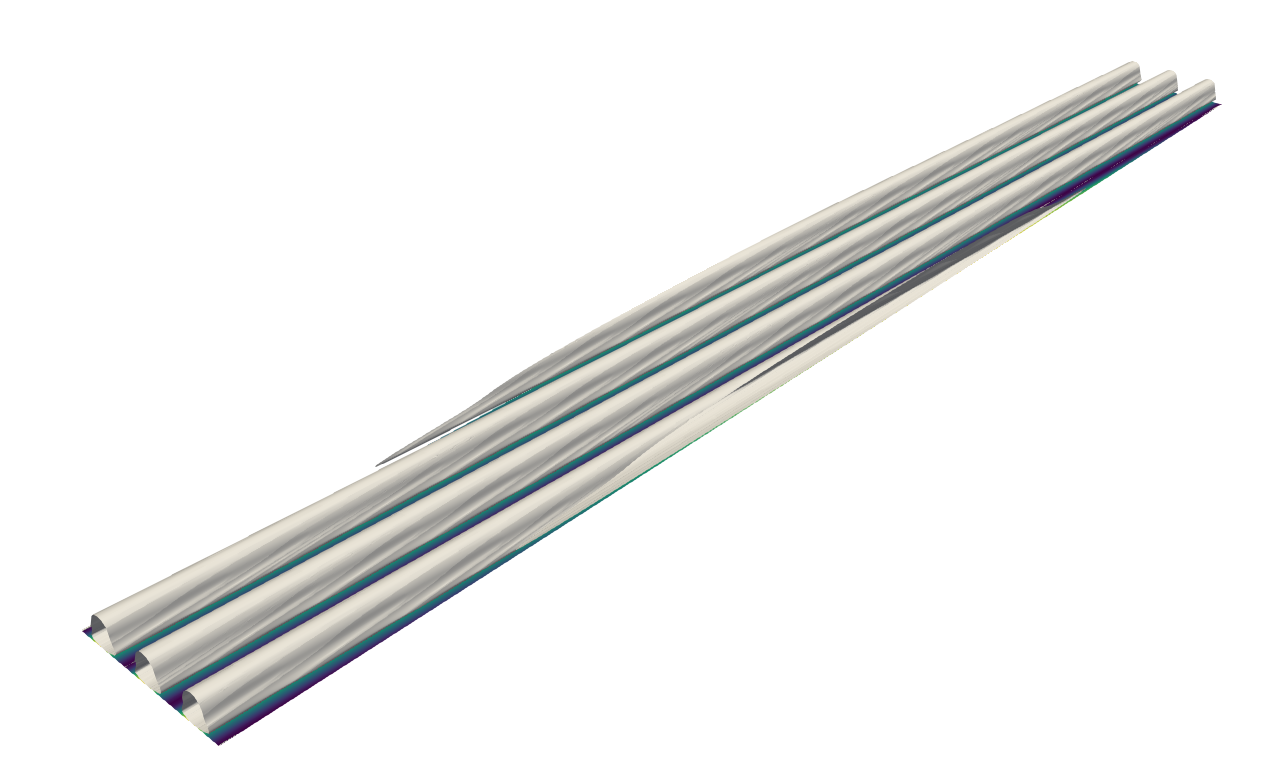} \\
    \includegraphics[trim={3cm 1cm 2cm 2cm},clip,width=0.65\textwidth,height=0.2\textheight]{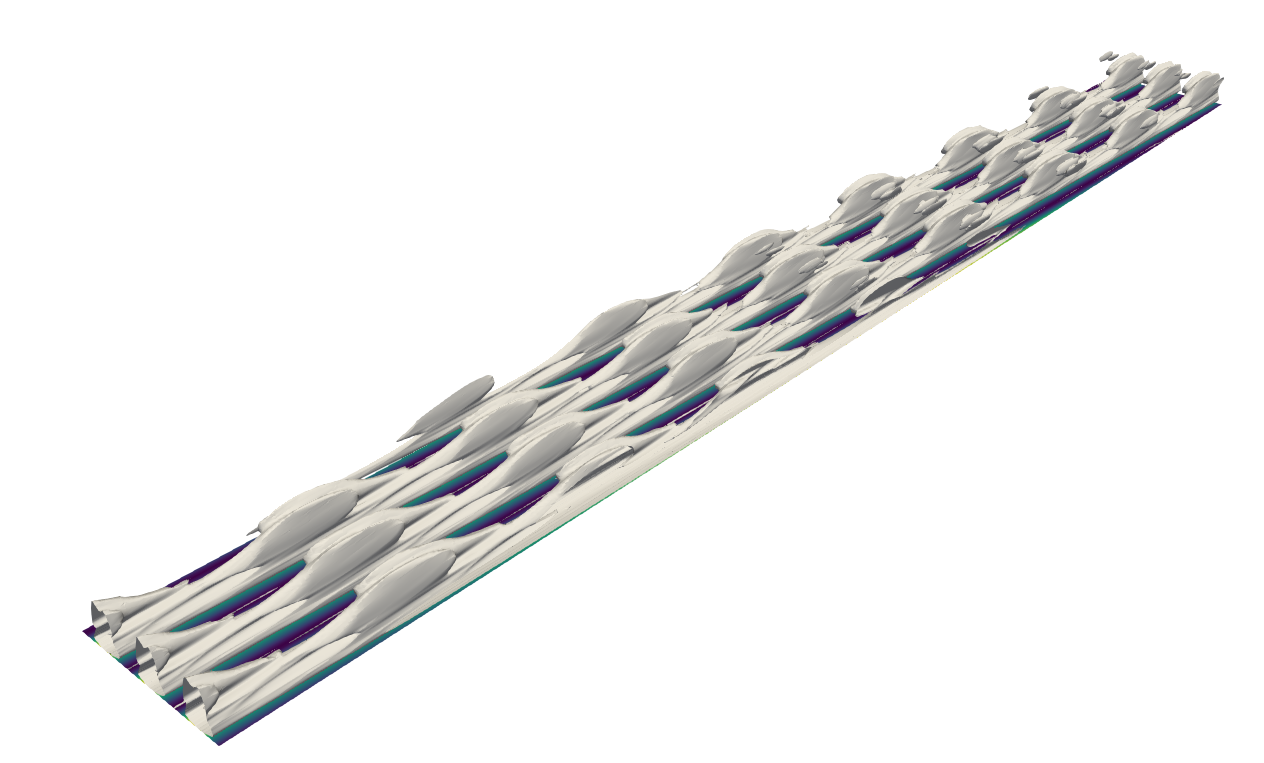} \\
    \includegraphics[trim={3cm 1cm 2cm 2cm},clip,width=0.65\textwidth, height=0.2\textheight]{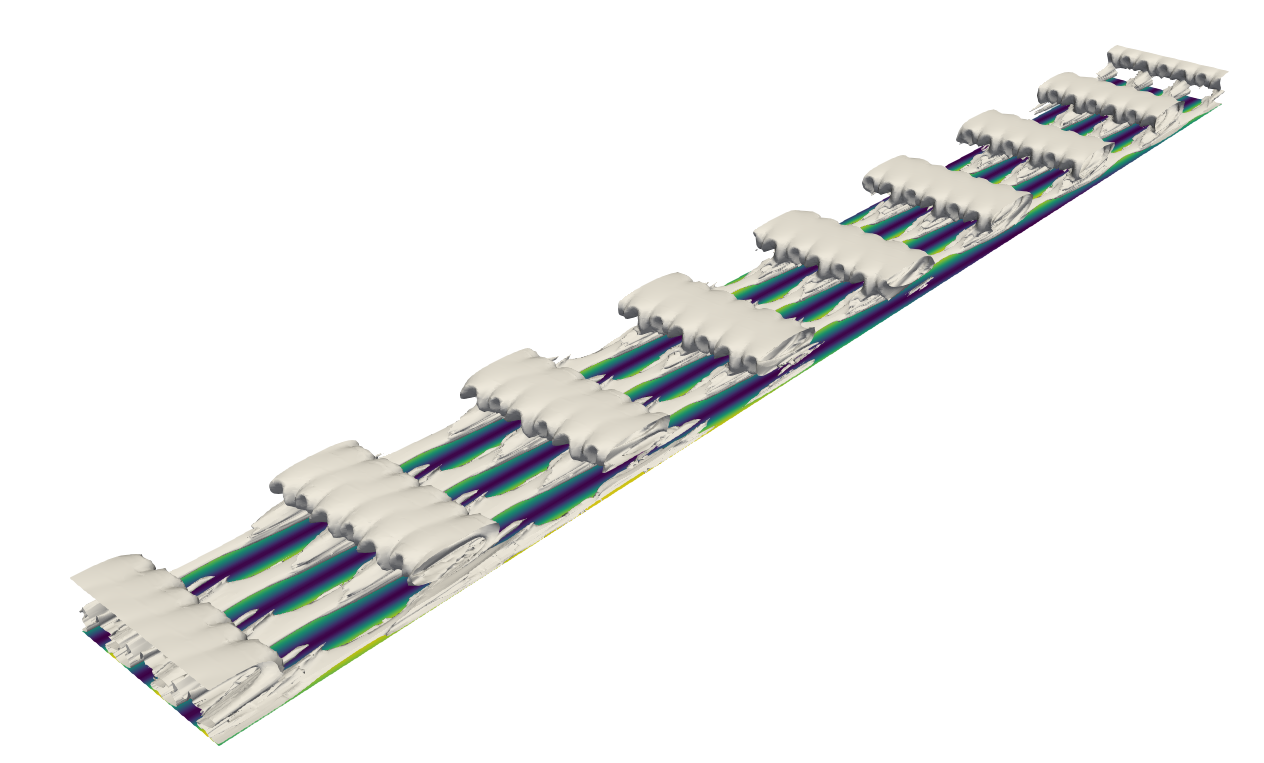} \\
    \includegraphics[trim={3cm 1cm 2cm 2cm},clip,width=0.65\textwidth, height=0.2\textheight]{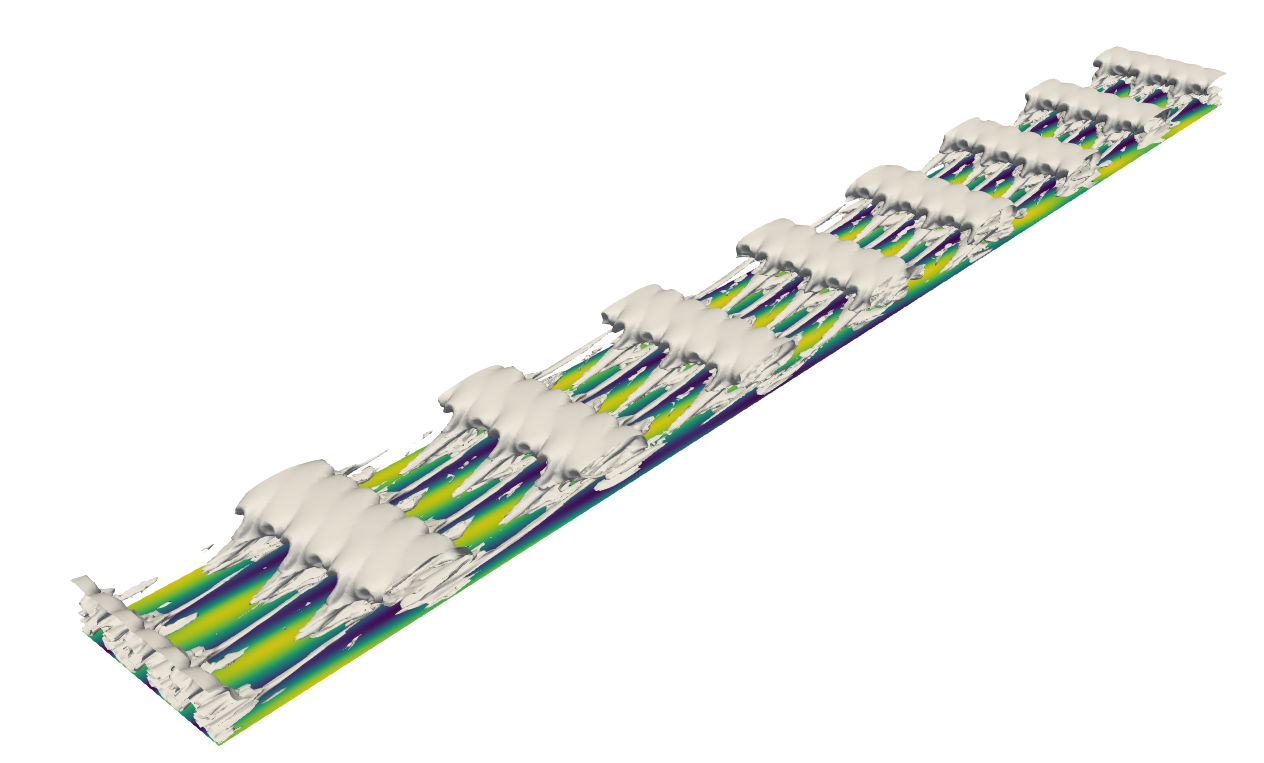} \\
    \includegraphics[trim={0cm 0cm 0cm 0cm},clip,width=0.65\textwidth, height=0.2\textheight]{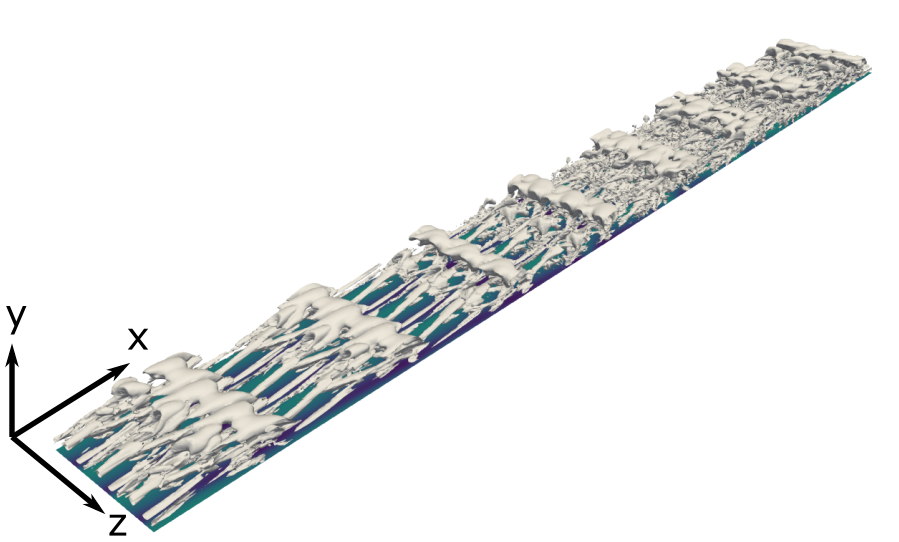} \\
    
    \caption{Snapshots of the flow at   $T=900$, $T=1000$, $T=1050$, $T=1060$, and $T=1080$ (from top to bottom). Isosurfaces of the $\lambda_2$-criterion, $\lambda_2=-10^{-5}, -10^{-4}, 0.25$, respectively, and contours of the streamwise velocity at the wall. The flow is from bottom to top, and left to right. For the sake of clarity, only half the channel is shown.}
    \label{fig:SnapshotsCF3D}
\end{figure}

At first, exponential growth of the disturbance kinetic energy is observed in figure \ref{fig:KeReTauCF}. Notice that the slope of the curve corresponds to twice the growth rate $\omega_r = 0.0033$ retrieved from  linear stability analysis. After $t=600$, the kinetic energy starts saturating at $E \approx 1.8\times 10^{-3}$. This saturation most likely arises from stabilising nonlinear effects and is also observed in a CF induced transition for swept flows. 
As shown in the first snapshot of figure \ref{fig:SnapshotsCF3D},
besides slight deformations near the walls due to the boundary conditions, the shape of the quasi-stationary ($\omega_i = 0.0225$) cross-flow vortices does not evolve much during this phase.

\begin{figure}
    \centering
    \includegraphics[width=0.45\textwidth]{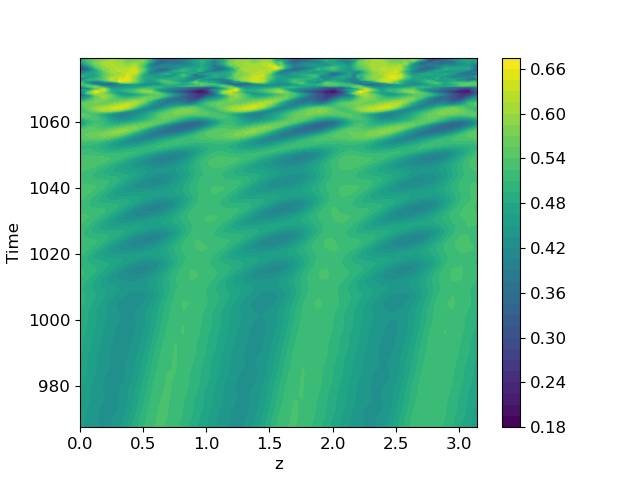}
    \includegraphics[width=0.45\textwidth]{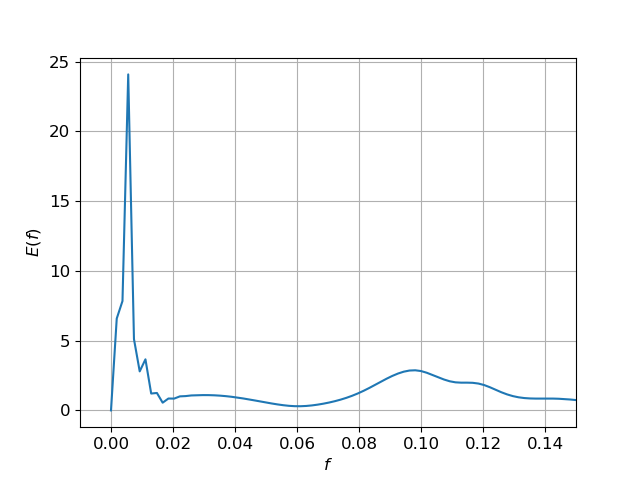}
    \caption{Left: Spatio-temporal evolution of the streamwise (left column) and spanwise (right column) components of the velocity at $x=0$ and $y=-0.85$. Right: Energy temporal spectrum of the transition to turbulence.}
    \label{fig:SpatioTemporalCF}
\end{figure}
In the meantime, it can be seen from figure \ref{fig:FourierCF} that the mode $(2,0)$ also experiences strong energy growth. Since the mode is linearly stable and is not expected to undergo strong transient growth, only non-linear mechanisms could explain this growth. Nonlinear forcing of the mode $(2,0)$ by the initial perturbation is possible as it can be observed in an oblique waves transition scenario for a boundary layer \citep{schmid_analysis_2014}. Usually, the $(2,0)$ mode is damped and not very receptive to forcing, meaning it is not relevant in the transition scenario. For 3D flows, \citet{el_hady_1989} demonstrated that resonant wave triads play an important role in the stability due to the large number of interaction between the possible instabilities. 

At $t = 1000$, the top of the vortices start oscillating in the streamwise direction (see bottom of figure \ref{fig:SnapshotsCF3D}). This can also be seen on the spatiotemporal plot of figure \ref{fig:SpatioTemporalCF}, where one can notice the much higher frequency of the secondary instability. This secondary instability can be traced back to the Fourier mode $(7,0)$ in figure \ref{fig:FourierCF}. Secondary growth rate can be extracted from the slope of its time evolution and is equal to $\sigma_r \approx 0.084$. The frequency associated to secondary instability can be obtained from the Fourier spectrum of figure \ref{fig:SpatioTemporalCF} and is equal to $\sigma_i \approx 0.095$.

\begin{figure}
    \centering
    \includegraphics[trim={0 9cm 0 9cm},clip,width=1.\textwidth]{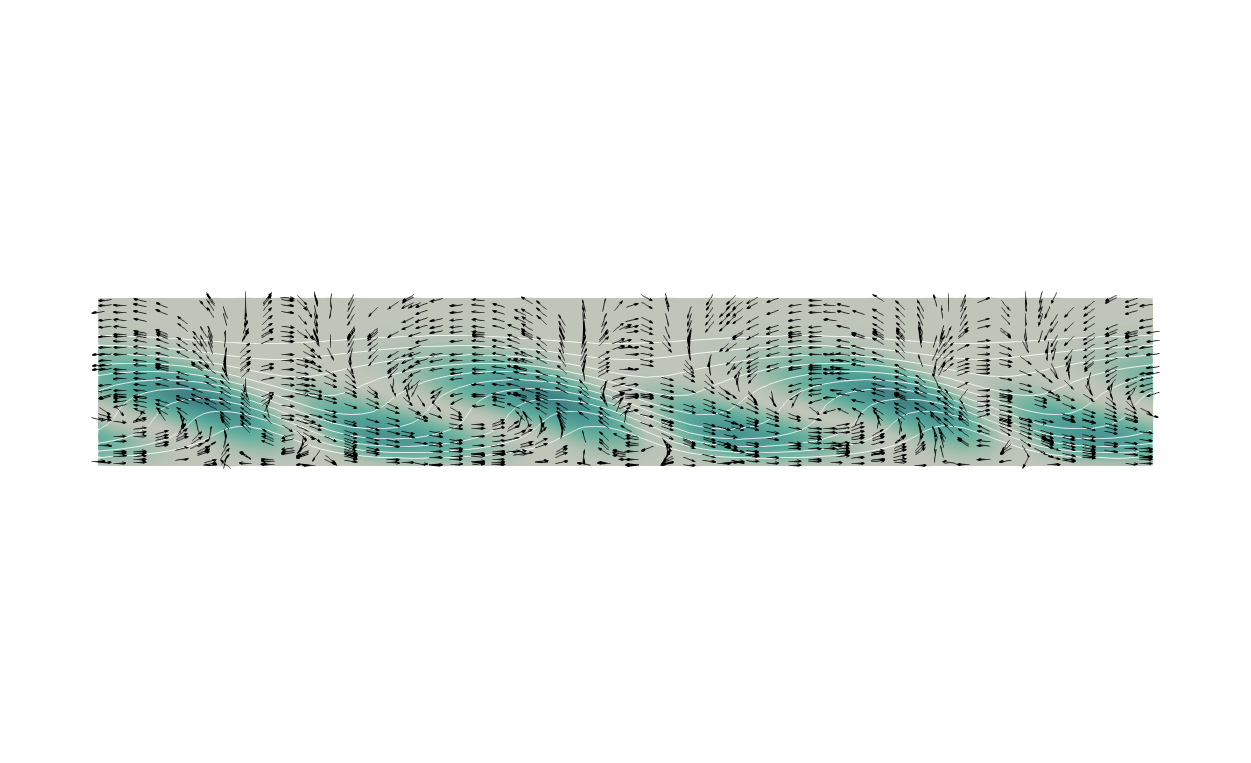} \\
    \includegraphics[trim={0 9cm 0 9cm},clip,width=1.\textwidth]{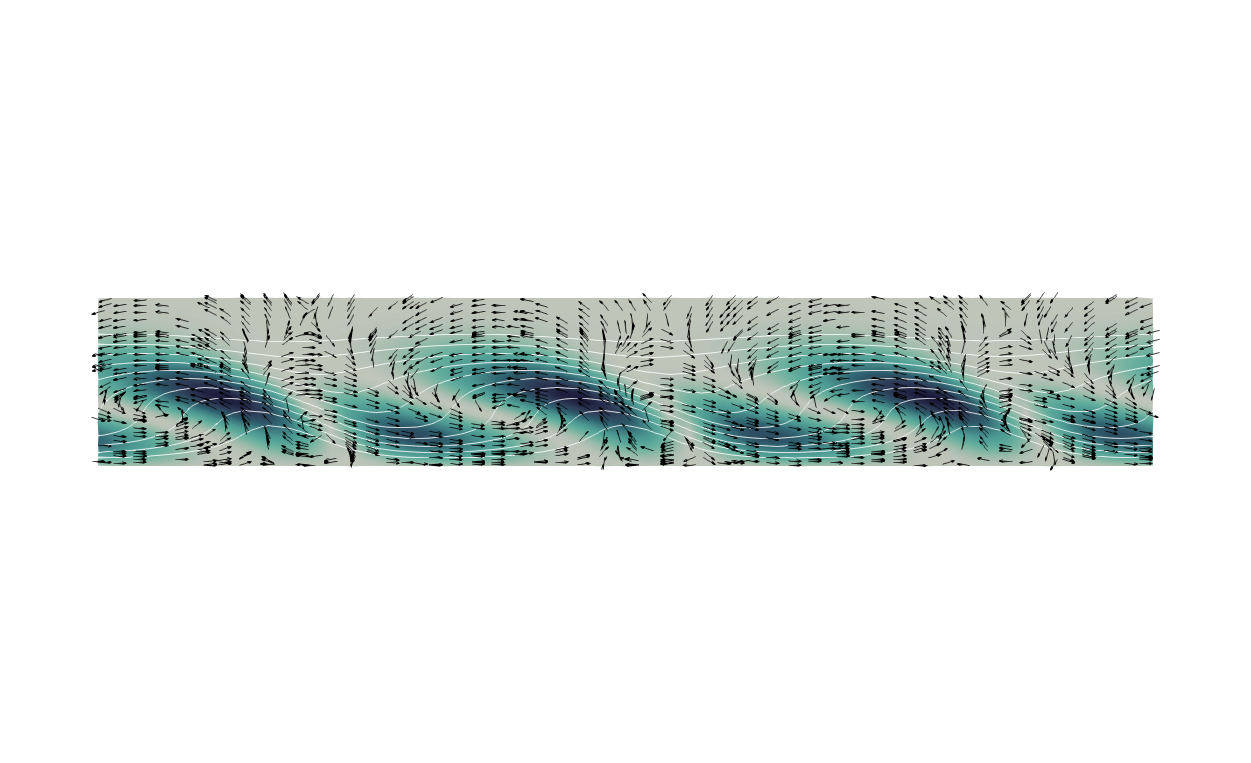} \\
    \includegraphics[trim={0 9cm 0 9cm},clip,width=1.\textwidth]{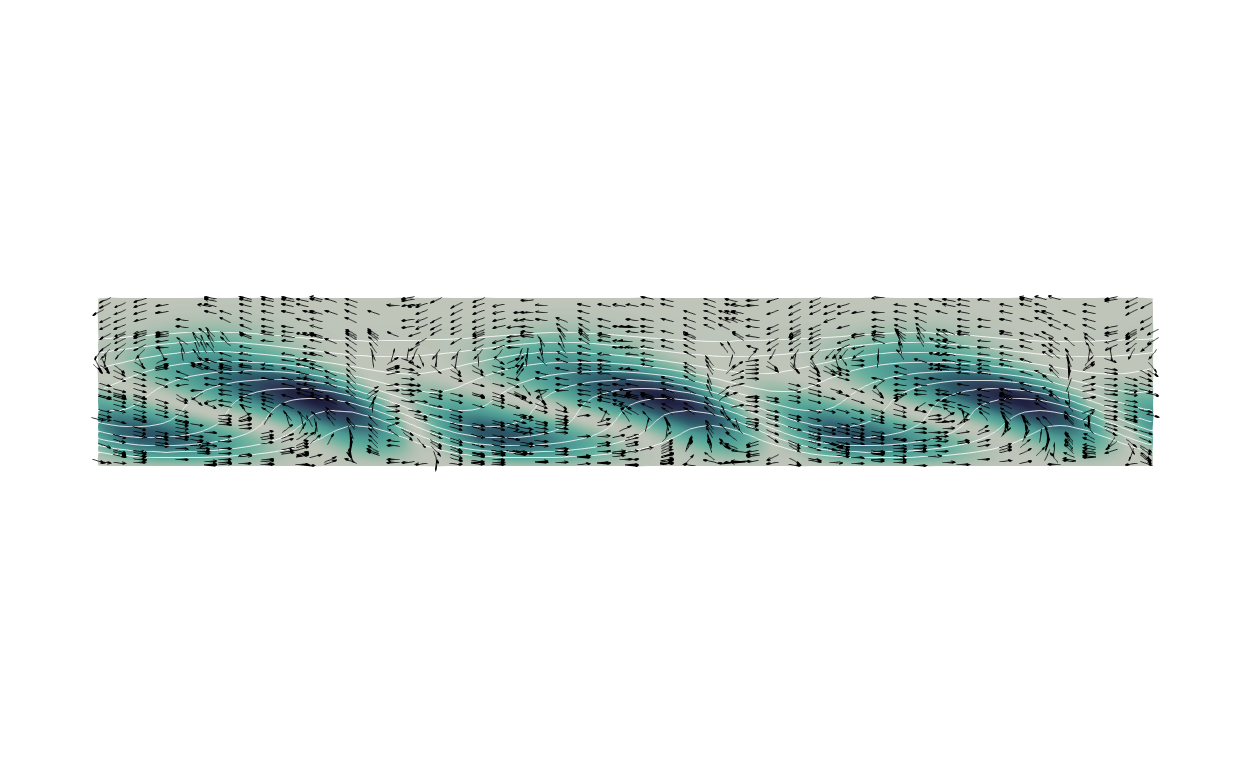} \\
    \includegraphics[trim={0 9cm 0 9cm},clip,width=1.\textwidth]{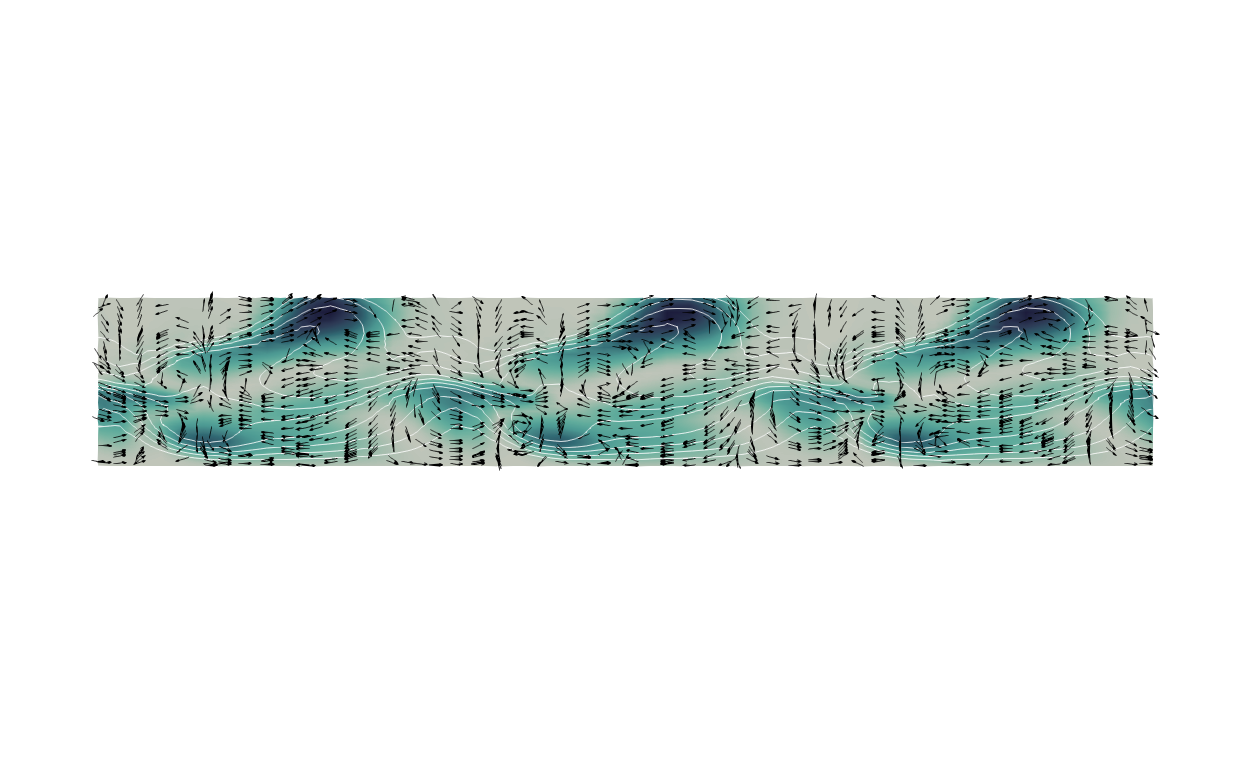} \\
    \includegraphics[trim={0 4cm 0 5cm},clip,width=1.\textwidth]{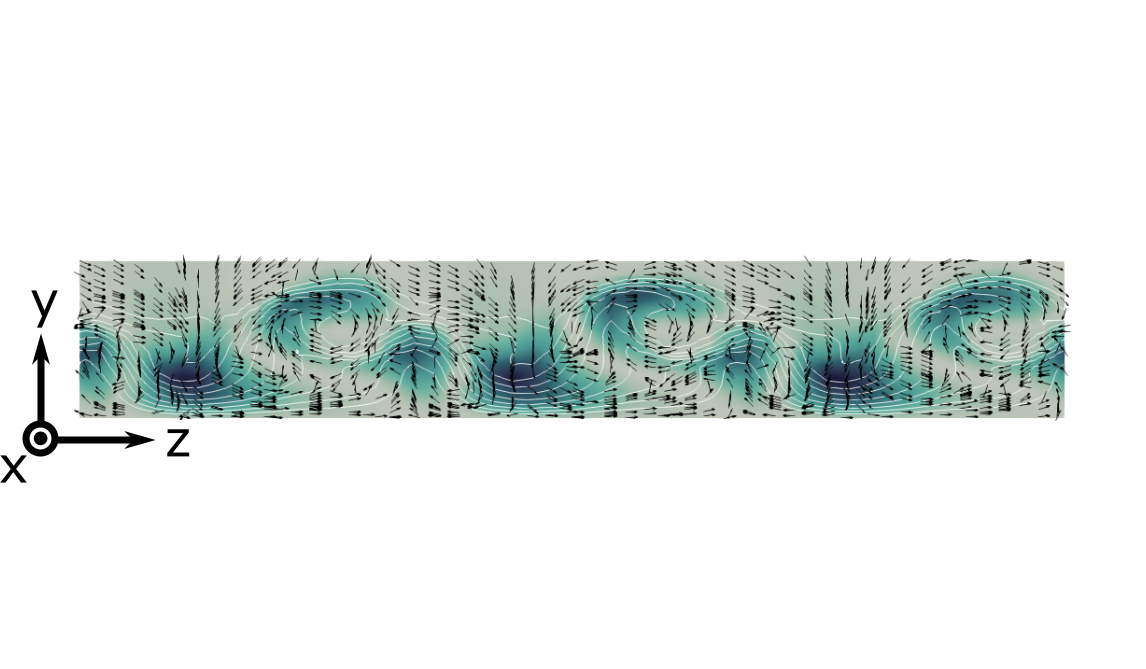} \\
    
    \caption{Cross-flow dynamics at different time T. From top to bottom: $T=900$, $T=1000$, $T=1050$, $T=1060$ and $T=1080$. Slice is taken at the streamwise position $x_s = 15.71$ and only $y \in [-1,-0.7]$ is represented. Isocontours of the $\lambda_2$-criterion [$-1\times 10^{-4} \longleftrightarrow -1\times 10^{-2}$]. Contour plot of the $u$-component of the disturbance velocity. Quiver plot depicts the $(v-w)$ crossflow of the perturbation. White contours represent the magnitude of the velocity $\left[-5\times 10^{-2}, -3\times 10^{-1}; -5\times 10^{-2}\right]$.}
    \label{fig:SnapshotsCFPlane}
\end{figure}

At $t \approx 1050$, spanwise independent coherent structures appear as shown in figure \ref{fig:SnapshotsCFPlane}. These structures appear on top of the cross-flow vortices and seem to be tilted. The bottom snapshot of figure \ref{fig:SnapshotsCFPlane} is particularly interesting: on top of the wave-like pattern of the initial disturbance, secondary vortices can be observed. These also appear in the near-wall region between two CF vortices. These vortices are most likely a consequence of the slip boundary conditions: as the fluid is accelerated in the spanwise direction near the walls, strong spanwise shear layers, which are susceptible to further destabilise, are created. Energy is transferred from the CF vortices to the secondary perturbation. Distortion of the base flow, in both streamwise and spanwise directions, is also highly likely.    

The remaining part of the laminar-turbulent transition is rather complex as several instability mechanisms get intertwined. In the fourth snapshot of figure \ref{fig:SnapshotsCF3D}, streaks can be seen developing in the near wall region, causing the energy growth of the $(0,1)$ Fourier mode in figure \ref{fig:FourierCF}. In figure \ref{fig:SnapshotsCFPlane}, the vortices previously described have combined to form vortex quadrupoles which create strong recirculation in the flow. Ultimately, the breakdown to turbulence seems to be related to the displacement of high-velocity fluid in the upper part of the channel (fourth snapshot of figure \ref{fig:SnapshotsCFPlane}) towards lower-velocity regions near the wall (last snapshot of figure \ref{fig:SnapshotsCFPlane}). In the vicinity of these regions turbulent wedges originate (see figure \ref{fig:SpatioTemporalCF}) which quickly propagate to the whole channel.

\subsection{Secondary stability analysis of cross-flow vortices}

Due to the presence of several strong shear layers in the primary cross-flow vortices, these are highly likely to destabilise. The observation of the transition scenario also strongly hints at the presence of modal secondary instability mechanisms. In the case of swept flows, numerous efforts on the secondary instability of cross-flow vortices have been realised, albeit, to the author's knowledge, always for boundary flows with no-slip boundary conditions. While a short introduction to secondary instability of cross flow vortices is produced in the following, the reader is referred to \citet{saric_stability_2003} for a full review.

One of the first key contribution to this issue has been the work of \citet{malik_secondary_1999} which classified the secondary modes into two main families. Precisely,  \textit{type-I} unstable modes are linked to the velocity gradient of the streamwise component of the velocity in the spanwise direction, while  \textit{type-II} modes originate from gradients in the wall-normal direction. \textit{Type-I} modes are located in the outer part of the primary vortices while \textit{type-II} are situated on top of them. \textit{Type-II} also tend to have a higher frequency than \textit{type-I} modes. A third type of mode was identified in studies by \citet{fischer_primary_1991} and \citet{janke_secondary_2000} through Floquet analysis of the cross-flow vortices. This low-frequency \textit{type-III} mode is linked to nonlinear interactions between stationary and travelling primary modes. These three families of modes have been also experimentally retrieved by \citet{kawakami_crossflow} and \citet{white_secondary_2005}. Later, using DNS, \citet{wassermann_mechanisms_2002,wassermann_transition_2003} showed that co-rotating helicoidal structures superimposed on the upwelling region of the primary vortices were characteristics of the \textit{type-I} mode. These structures would be convected downwards if the unsteady forcing was switched off, thus confirming the convective nature of the secondary instability as already suggested by \citet{kawakami_crossflow} and \citet{koch_stability}.

\begin{figure}
    \centering
    \includegraphics[width=1.\textwidth]{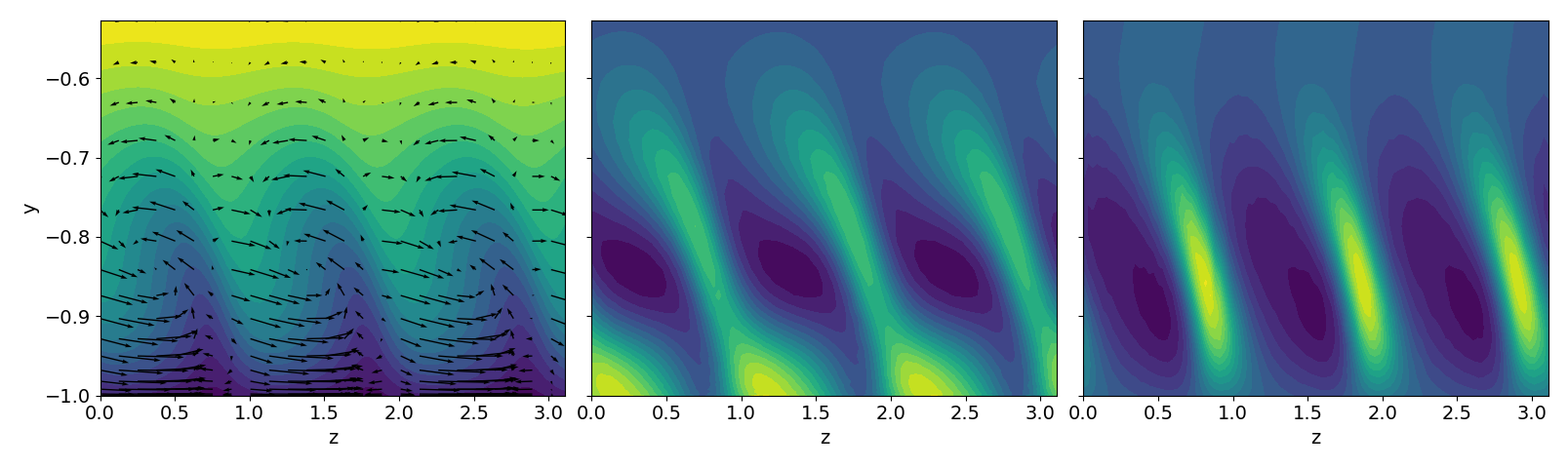}
    \caption{Base flow for the secondary stability analysis. The snapshot is extracted from the DNS at time $t \approx 900$, position $x_{SLST} = 15.71$ and is shown in the vortex-oriented reference frame. Left: contour plot of the velocity component $U$. Quiver plot of the cross-flow perturbation $(v-w)$. Constant spanwise $W_0$ has been substracted from $w$ in order to observe the vortical structures. Middle: wall-normal gradient of the streamwise component of velocity $\partial_{y} u^0_v$. Right: spanwise gradient of the streamwise component of velocity $\partial_{z_v} u^0_v$.  }
    \label{fig:BFDNS}
\end{figure}

\citet{bonfigli_secondary_2007} also studied the development of the secondary instabilities of cross-flow vortices through the combined use of a spatial DNS and Secondary Linear Stability Theory (SLST) based on Floquet theory applied on a saturated primary flow \citep{malik_crossflow_1994,janke_secondary_2000,koch_secondary, koch_stability}. Growth rates and structures of the \textit{type-I} and \textit{type-III} modes were retrieved and good agreement was shown between the two techniques. SLST predicted the instability of \textit{type-II} modes but these were not observed in the DNS. In the particular case of travelling vortices, \textit{type-III} modes were not reported as unstable, which was expected since this mode arises from the generation of spanwise modulation induced by the stationary primary vortices. Another key conclusion of this study is the nature of the mechanism behind the secondary instability: both \textit{type-I} and \textit{type-II} are found to be related to Kelvin-Helmoltz instabilities.
In all cases, the primary flow on which SLST is realised is paramount and its acquisition is not straightforward as nonlinearities lead to an important distortion of the base flow. The shape assumption does not hold anymore and Herbert's secondary stability theory cannot be directly applied. 

The primary flow can be retrieved by using parabolised stability equations as in \citet{malik_secondary_1999} and \citet{janke_secondary_2000}, or directly extracted from a saturated DNS \citep{messing_2004, bonfigli_secondary_2007}. Recently, \citet{groot_2018} based their secondary stability analysis on an experimentally acquired primary flow. In any case, the extraction process is quite complex and several issues arise. In swept boundary layers, the orientation of the vortex axis of the cross-flow vortices evolves with the streamwise direction. Consequently, secondary linear stability depends on the position $x=x_{SLST}$ where the primary flow is extracted. This is not the case in our configuration, as a consequence of the choice of a temporal DNS. 

As in \citet{bonfigli_secondary_2007} and several others, we introduce a vortex-oriented system $(x_v,y,z_v)$ aligned with the vortex axis of the primary disturbance and moving with it. In this frame, the primary flow $\mathbf{Q}_v^0(y,z_v)$ is streamwise independent and periodic in the spanwise direction. In the framework of Floquet theory, it is Fourier expanded in the spanwise direction and takes the following modal form:
\begin{equation}
    \mathbf{Q}_v^0(y,z_v) = \sum_{n=-\infty}^{+\infty} \mathbf{\Hat{q}}^0_{v,n}(y)e^{in\beta_vz_v}
\end{equation}
with $\beta_v =\sqrt{\alpha_0^2 + \beta_0^2} \approx \beta_0$ the effective wavelength in the spanwise direction $z_v$. In this framework, the continuity equation reduces to:
\begin{equation}
    \partial_y v^0_{v,n} + in\beta_vw^0_{v,n} = 0, \qquad \forall n,
\end{equation}

directly relating the two cross-flow components of the velocity. Thus, a choice must be made between enforcing continuity and extracting both components from the DNS. Ultimately, this might have a non negligible impact on the secondary growth rates found from the SLST \citep{malik_crossflow_1994, bonfigli_secondary_2007}. 

To overcome this and other issues of  Floquet analysis, here we use a bi-local stability \citep{tatsumi_yoshimura_1990} approach. Several reasons motivate this choice: first and foremost, two-dimensional local stability theory encompasses Floquet stability theory. Also, as previously seen, Fourier decomposition in the spanwise direction of the primary flow leads to some difficulties with the continuity equation. In a bilocal framework, this decomposition is not necessary and these issues do not arise. Ultimately, bilocal stability analysis allows to take into account also collective instability coupling multiple cross-flow vortices, while  other studies only considered a single pattern of the primary flow for their secondary stability analyses. 

For carrying out the bilocal secondary stability analysis, the flow is decomposed into the primary state $\mathbf{Q}_v^0(y,z_v)$ and a secondary perturbation $\mathbf{q}_1(x_v,y,z_v,t)$. Furthermore, the secondary perturbation can be expanded in the following way:

\begin{equation}
    \mathbf{q}_1(x_v,y,z_v,t) = \Tilde{\mathbf{q}}(y,z_v)e^{i\alpha_v x_v - \sigma t}   
    \label{BilocalExpansion}
\end{equation}

where $\alpha_v$ and $\sigma$ are complex numbers representing respectively the streamwise wavelength and the pulsation in the vortex-oriented reference frame. Both temporal  \citep{malik_secondary_1999,koch_secondary, wassermann_mechanisms_2002} and spatial \citep{janke_secondary_2000, bonfigli_secondary_2007} approaches have been previously considered. \citet{koch_secondary} found a set of equations relating both frameworks through a generalisation of the Gaster transformation to 3D flows. For channel flows, a natural choice is to consider a temporal framework. As previously, Navier-Stokes are linearised around this new base flow. Stability equations for two-dimensional problems can be found for example in \citet{loiseau_dynamics_2014}.

The primary flow in the vortex oriented referential is shown in figure \ref{fig:BFDNS}. It was extracted from the DNS at time $t=900$ and $x_{SLST}=15.71$. This flow field can be compared with the flow visualisations from the experiments of \citet{serpieri_cross-flow_2018} (see figure 4.14) and with the primary flow resulting from the DNS of \citet{bonfigli_secondary_2007} (see figure 7). The water-wave shape of the streamwise velocity component, characteristic of the cross-flow vortices, is clearly visible. However, the spatial extension of the tip of the wave is smaller than that observed for  swept boundary flows, likely because the initial spanwise velocity $W_0$ is weaker. The instability mechanism is similar to that described in \citet{serpieri_three-dimensional_2016}: cross-flow vortices generate the circulation of low-momentum flow towards high-momentum regions higher up in the channel and conversely. In the present case, the wall-normal shear is maximum near the walls and not in the vortex core due to the slip boundary conditions. The spanwise velocity gradient reaches a maximum in the low-momentum upwelling region on the outer part of the vortex.

\begin{figure}
    \centering
    \includegraphics[width=0.6\textwidth]{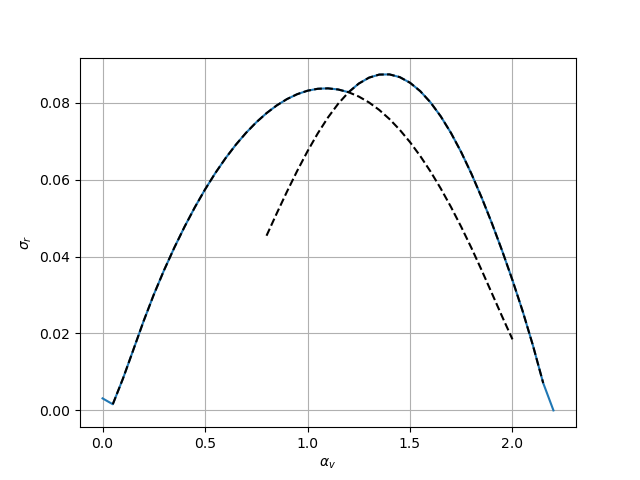}
    \caption{Growth rate $\sigma_r$ of the secondary most unstable modes of CF vortices as a function of the effective streamwise wavenumber $\alpha_v$ in the vortex-oriented frame.}
    \label{fig:SecondGrowth}
\end{figure}

Bilocal stability analysis of the previously described primary flow is performed. The evolution of the temporal secondary amplification rates in the vortex-oriented frame as a function of the effective streamwise wavelength is shown in figure \ref{fig:SecondGrowth}. The maximum growth rate is equal to $\sigma_r = 0.084$ for $\alpha_v=1.4$. Both the growth rate and the effective streamwise wavenumber are in good agreement with the ones observed in the DNS. A direct comparison between the characteristics of the secondary instability obtained from the SLST and those extracted from the DNS is provided in table \ref{tab:SLST}. 

\begin{table}
  \begin{center}
\def~{\hphantom{0}}
  \begin{tabular}{lccc}
                 & $\alpha_v$ & $\sigma_r$   &   $\sigma_i^0$ \\[3pt]
        DNS      &   1.4   & 0.082 & 0.095 \\
        2D SLST  &   1.39  & 0.087 & 0.088
  \end{tabular}
  \caption{Comparison between secondary instability characteristics obtained from the DNS and from the most unstable mode of the secondary stability analysis. }
  \label{tab:SLST}
  \end{center}
\end{table}

All the results from the SLST are obtained in the moving vortex-oriented frame. Values of streamwise wavelengths and frequencies in the laboratory frame can be retrieved by expanding equation (\ref{BilocalExpansion}):

\begin{align}
\sigma_i^{0} &= \sigma_{i,v} - \alpha_v c \\
\alpha_1^{0} &= -\frac{\alpha_v \beta_0}{k} \\
\beta_1^{0} &= \frac{\alpha_v \alpha_0}{k}
\end{align}

where $k = \sqrt{\alpha_0^2 + \beta_0^2}$ and $c$ is the velocity of the CF vortices. 

\begin{figure}
    \centering
    \includegraphics[width=0.45\linewidth]{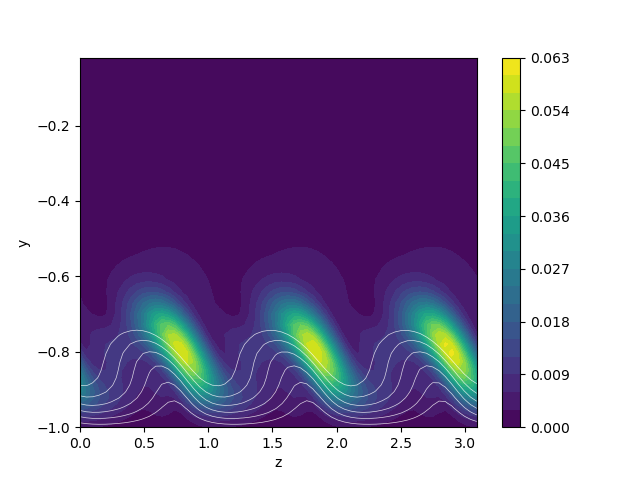}
    \includegraphics[width=0.45\linewidth]{Images/Mode1SecondStabAbs.png}
    \includegraphics[width=0.45\linewidth]{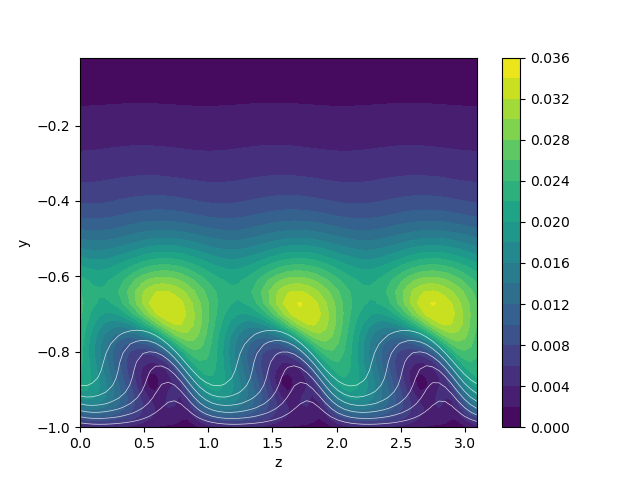}
    \caption{Absolute value of the eigenfunctions of the most unstable secondary mode of CF vortices for $Re=12000$ and (top left) $\alpha_v=1.1$, (top right), $\alpha_v=1.4$ (bottom) $\alpha_v=0.2$  In white: isocontours of the streamwise component of the primary flow ($[0.1 \longleftrightarrow 0.35]$).}
    \label{fig:ModesSecond}
\end{figure}

Figure \ref{fig:SecondGrowth} also displays a competition between two types of modes. The eigenfunctions for these two modes can be retrieved in figure \ref{fig:ModesSecond} and are found to be almost identical. The amplitude of the streamwise perturbation is strongly localised on the upwelling of the vortex, a representative feature of \textit{type-I} modes. Among the several unstable modes, no \textit{type-II} nor \textit{type-III} structures could be found. This is not unexpected: \textit{type-III} modes require interactions with unstable stationary cross-flow vortices. \textit{Type-II} modes stem from the instability of wall-normal velocity gradients but these are sensibly weaker than their counterparts in swept flows. This could be explained by the combined effect of the absence of an inflection point in the spanwise velocity profile $W_0$ and of the slip boundary conditions. Some insight can be gained on the nature of the instability by looking at the eigenfunctions for small wavelengths. An example for $\alpha_v=0.2$ is displayed in the bottom of figure \ref{fig:ModesSecond} and shows a superposition of the primary flow almost undisturbed and, on top of it, a slightly deformed "cat's eye" pattern, characteristic of a Kelvin-Helmholtz instability.

\begin{figure}
    \centering
    \includegraphics[trim={0 8cm 0 5cm},clip, width=0.45\textwidth]{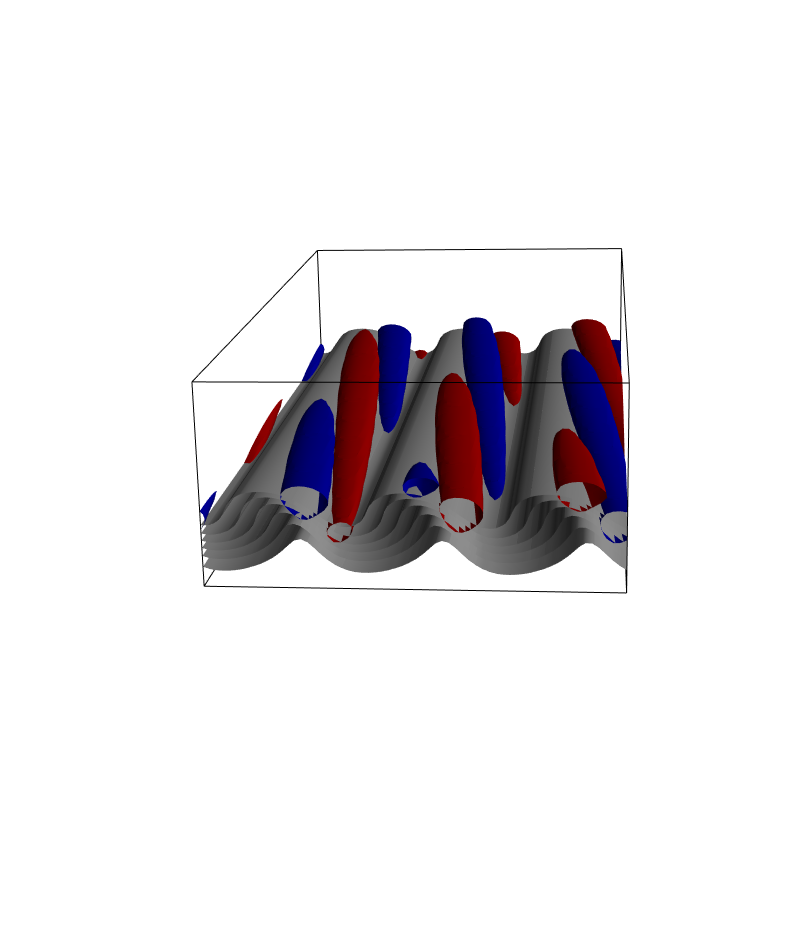}
    \includegraphics[trim={0 8cm 0 5cm},clip, width=0.45\textwidth]{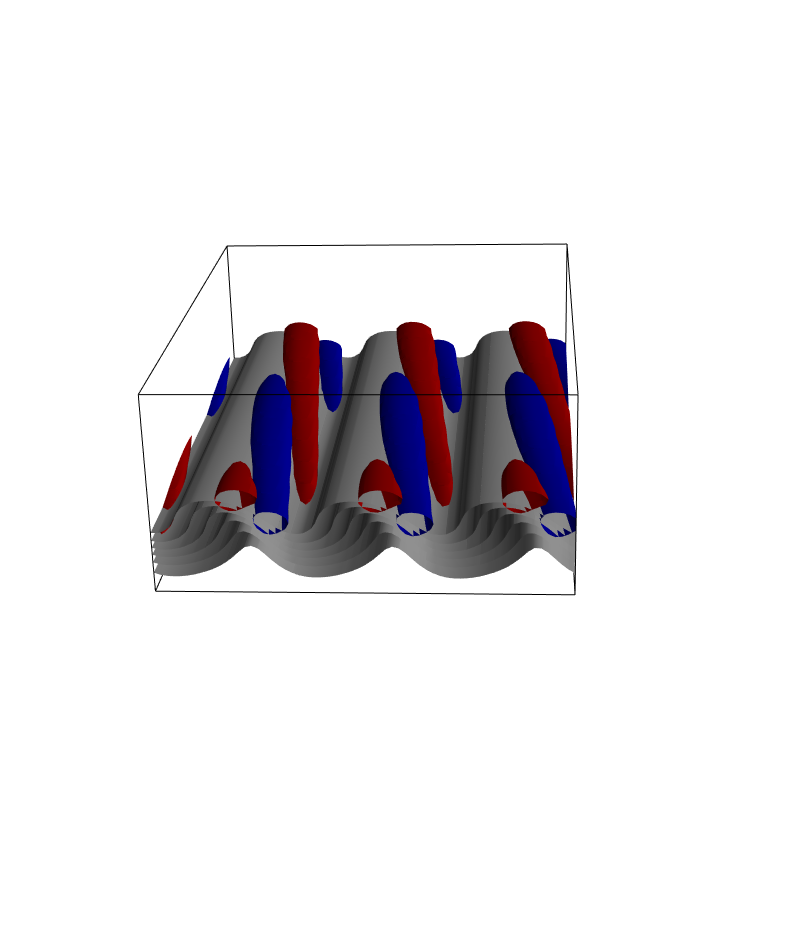}
    \caption{In red and blue: streamwise velocity component for $\alpha_v=1.1$ (left) and $\alpha_v=1.4$ (right), 
for $\tilde{u}\pm 0.75 u_1^{max}$, respectively.  In grey: isocontours of the streamwise component of the primary flow ($[0.1 \longleftrightarrow 0.35]$).}
    \label{fig:ModesSecond3d}
\end{figure}

The two modes have been fully reconstructed from equation (\ref{BilocalExpansion}) and superposed on the primary flow in figure \ref{fig:ModesSecond3d}. The difference between the two modes might result from the collective behaviour of the secondary instability. For $\alpha_v=1.4$, the secondary perturbations are in phase on each wave-like pattern of the primary flow. Thus, a single CF vortice would have been sufficient for determining the secondary instability. On the contrary, in the $\alpha_v=1.1$ case, a phase shift appears between the multiple iterations of the secondary perturbations, thus coupling multiple CF vortices together. Finally, as observed in the DNS, the synchronised perturbation is the most unstable.

\section{Conclusions and outlook}\label{sec:Conclusions}

In this paper, the influence of roughness anisotropy on the laminar-turbulent transitions of superhydrophobic surfaces modelled with homogenised, generalised slip-like boundary conditions is investigated. Following \citet{pralits_stability_2017}, anisotropy of the surface is taken into account through the use of a mobility tensor generalising the Navier slip condition. This set of homogenised boundary conditions leads to the presence of a cross-flow  component in the base flow. Despite its small magnitude and its lack of an inflection point, this cross-flow component deeply affects the linear stability of the flow. In particular, a new instability region appears for small streamwise and large spanwise wavenumbers. The underlying instability mechanism was found to be similar to the cross-flow instability observed in swept flows. TS waves could also be retrieved but the most unstable ones are 3D due to the asymmetry of the neutral curve, induced by the cross-flow. Transient growth has been also considered, showing that the optimal perturbation has the form of oblique streaks with
non-zero streamwise wavenumber. In the case of a constant spanwise velocity, a generalisation of the lift-up effect can be derived. 

Direct numerical simulations have been performed in two distinct flow cases showing different unstable modes. In the TS-wave initiated transition, instead of the usual subharmonic bifurcation, streaky structures with large spanwise wavenumber have been found, which ultimately destabilise and broke down into turbulence. The second scenario, initiated with a cross-flow related unstable mode, depicts a highly nonlinear transition, with the flow rapidly saturating towards "half-mushroom" structures as seen in \citet{malik_crossflow_1994}, before ultimately transitioning to turbulence. In both transition scenarios, secondary stability analysis has been performed. In particular, Floquet analysis has been used for the TS-wave primary instability, while BiGlobal stability analysis has been used for the cross-flow mode transition. In both cases, it has been shown that the underlying instability mechanisms are linked to the presence of the cross flow, thus stressing the importance of modelling anisotropy when studying superhydrophobic surfaces through homogenised boundary conditions. It has been also shown that coherent structures such as the streamwise vortices or the Kelvin-Helmoltz rollers are retrieved by means of secondary stability analysis, and they have a non negligible influence on the transition scenarios. 
For the transition induced by cross-flow vortices, the importance of coupling effects has been established. 

It remains to be verified if a different riblet orientation, chosen in our case to be $45^o$ for the sake of simplicity, would yield significantly different results. A more refined modelling of the superhydrophobic surfaces could also be considered and eventually lead to an even more accurate characterisation of the laminar-turbulent transition. Among the possible improvements, the homogenised boundary condition could, for example, be replaced with a patterned slip-like/no-slip boundary condition. Also possible, but highly non trivial from both analytical and computational perspectives, the deformation of the gas-liquid interface could be considered. Future works will aim at investigating these points. 

\appendix
\section{Lift-up in a 3D flow}\label{appendix:liftup}

The lift-up effect has initially been described by \citet{ellingsen_stability_1975}. Its generalisation from 3D flows remains unclear. Thus, following \citet{ellingsen_stability_1975} and \citet{brandt_lift-up_2014}, we consider a flow in a channel between two superhydrophobic walls. The resulting velocity profile $(U_0(y),0,W_0(y))$ is parallel and three-dimensional. The fluid is inviscid, incompressible and non-stratified. Perturbations are assumed with the following modal form $u(y) \exp{(i\beta z)}$, with a streamwise invariance and a spanwise wavelength $\beta$. Thus, the linearised momentum equations yield:

\begin{align}
    \frac{\partial u}{\partial t} + i\beta W_0 u + vU_0' &= 0 \\
    \frac{\partial v}{\partial t} + i\beta W_0 v &= 0 \\
    \frac{\partial w}{\partial t} + i\beta W_0 w + vW_0' &= 0
\end{align}

Introducing a streamfunction $\psi$ for the cross-stream components,

\begin{equation}
    v = \partial_z \psi; \qquad w = -\partial_y \psi 
\end{equation}

it is possible to obtain an equation on $\nabla^2_s \psi$ with $\nabla^2_s$ the two-dimensional laplacian in the $y-z$ plane:

\begin{equation}
    \frac{\partial}{\partial t} \nabla^2_s \psi = -i\beta W_0 \nabla^2_s \psi + i\beta W_0'w + \partial_y v W_0' + v W_0''
    \label{Lift}
\end{equation}

In the special case of a channel with two superhydrophobic walls, the spanwise component of the base flow $W_0$ becomes constant, and  Eq.(\ref{Lift}) reduces to: 
\begin{equation}
    \partial_t \nabla_s^2 \psi = -i\beta W_0 \nabla_s^2 \psi
    \label{CS}
\end{equation}

Now, unlike the usual lift-up effect, cross-stream components are now dependant of time: they oscillate with frequency $\beta W_0$. Since $v(t) \propto e^{-i\beta W_0 t}$ as suggested by Eq.(\ref{CS}), the linearised streamwise momentum equation can be integrated in time to obtain:

\begin{equation}
    u(t) = u_0 \cos{(\beta W_0t)} - v_0U't
\end{equation}

The modified lift-up effect induces, for early times, an oscillation of frequency $\beta W_0$ in time of the streamwise velocity of the perturbation. The classic algebraic growth is also retrieved. 

\section{Secondary stability equations}\label{app:secondary}

Secondary stability equations in their primitive variable formulation are quickly presented. Assuming a small perturbation $\mathbf{q}_1(\mathbf{x},t)$ from the base flow $\mathbf{U}_1$, linearised Navier-Stokes equations read:

\begin{multline}
    \partial_{t} u_1 + (U_0-c_x)\partial_{x}u_1 + (W_0-c_z)\partial_{z}u_1 + U'_0v_1 = -\partial_{x} p_1 + Re^{-1}\nabla^2 u_1 \\
    - A[u^{TS}_j \partial_j u_1 + u_1\partial_{x} u^{TS} + v_1\partial_{y} u^{TS}]
    \label{SecEq1}
\end{multline}

\begin{multline}
    \partial_{t} v_1 + (U_0-c_x)\partial_{x}v_1 + (W_0-c_z)\partial_{z}v_1 = -\partial_{y} p_1 + Re^{-1}\nabla^2 v_1 \\
    - A[u^{TS}_j \partial_j v_1 + u_1\partial_{x} v^{TS} + v_1\partial_{y} v^{TS}]
\end{multline}

\begin{multline}
    \partial_{t} w_1 + (U_0-c_x)\partial_{x}w_1 + (W_0-c_z)\partial_{z}w_1 = -\partial_{z} p_1 + Re^{-1}\nabla^2 w_1 \\
    - A[u^{TS}_j \partial_j w_1 + u_1\partial_{x} w^{TS} + v_1\partial_{y} w^{TS}]    
\end{multline}

\begin{equation}
    \partial_{x}u_1 + \partial_{y} v_1 + \partial_{z}w_1 = 0
    \label{SecEq2}
\end{equation}

Introducing (\ref{Floquet2}) in Eq. (\ref{SecEq1})-(\ref{SecEq2}) and rearranging sums, the perturbation equations for the $m$-th mode can be found:

\begin{multline}
    \sigma \Tilde{u}_m + i\alpha_m (U_0 - c_x)\Tilde{u}_m + i\beta_1 (W_0-c_z)\Tilde{u}_m + U'_0\Tilde{v}_m = -i\alpha_m\Tilde{p}_m + Re^{-1}(\mathcal{D}^2 - k_m^2) \Tilde{u}_m \\
    -A\{ [i\alpha_{m-1}u^{TS} + v^{TS}\mathcal{D} + i\beta_1 w^{TS}] \Tilde{u}_{m-1}
    + iku^{TS}\Tilde{u}_{m-1} + \mathcal{D}u^{TS}\Tilde{v}_{m-1}\} \\
    -A\{ [i\alpha_{m+1}(u^{TS})^* + (v^{TS})^*\mathcal{D} + i\beta_1 (w^{TS})^*] \Tilde{u}_{m+1}
    - ik(u^{TS})^*\Tilde{u}_{m+1} + \mathcal{D}(u^{TS})^*\Tilde{v}_{m+1} \}
\end{multline}
\begin{multline}
    \sigma \Tilde{v}_m + i\alpha_m (U_0 - c_x)\Tilde{v}_m + i\beta_1 (W_0-c_z)\Tilde{v}_m = -\mathcal{D}\Tilde{p}_m + Re^{-1}(\mathcal{D}^2 - k_m^2) \Tilde{v}_m \\
    -A\{ [i\alpha_{m-1}u^{TS} + v^{TS}\mathcal{D} + i\beta_1 w^{TS}] \Tilde{v}_{m-1}
    + ikv^{TS}\Tilde{u}_{m-1} + \mathcal{D}v^{TS}\Tilde{v}_{m-1}\} \\
    -A\{ [i\alpha_{m+1}(u^{TS})^* + (v^{TS})^*\mathcal{D} + i\beta_1 (w^{TS})^*] \Tilde{v}_{m+1}
    - ik(v^{TS})^*\Tilde{u}_{m+1} + \mathcal{D}(v^{TS})^*\Tilde{v}_{m+1} \}
\end{multline}
\begin{multline}
    \sigma \Tilde{w}_m + i\alpha_m (U_0 - c_x)\Tilde{w}_m + i\beta_1 (W_0-c_z)\Tilde{w}_m = -i\beta_1\Tilde{p}_m + Re^{-1}(\mathcal{D}^2 - k_m^2) \Tilde{w}_m \\
    -A\{ [i\alpha_{m-1}u^{TS} + v^{TS}\mathcal{D} + i\beta_1 w^{TS}] \Tilde{w}_{m-1}
    + ikw^{TS}\Tilde{u}_{m-1} + \mathcal{D}w^{TS}\Tilde{v}_{m-1}\} \\
    -A\{ [i\alpha_{m+1}(u^{TS})^* + (v^{TS})^*\mathcal{D} + i\beta_1 (w^{TS})^*] \Tilde{w}_{m+1}
    - ik(w^{TS})^*\Tilde{u}_{m+1} + \mathcal{D}(w^{TS})^*\Tilde{v}_{m+1} \}
\end{multline}
\begin{equation}
    i\alpha_m \Tilde{v}_m + \mathcal{D}\Tilde{v}_m + i\beta_1\Tilde{w}_m = 0
\end{equation}

with $\alpha_m = (m+\epsilon)k$, $k_m^2 = \alpha_m^2 + \beta_1^2$, $\mathcal{D} = \partial_{y_1}$ the partial derivative in the $y_1$ direction and $u^*$ denotes the complex conjugate of $u$. The boundary conditions being already linear, they are left unchanged and Eq.(\ref{BC}) also applies to the perturbation problem. 

\bibliography{main}
\end{document}